\documentclass[11pt]{article}
\usepackage[top=1in, right=0.8in, left=0.8in, bottom=1in]{geometry}
\usepackage{amsmath, amsfonts, amsthm, amssymb, enumerate, color, authblk, color, bm, graphicx, multirow, comment, url, xspace, algorithm2e, natbib, titlesec, hyperref, ulem, lscape,float}
\usepackage{hhline}
\usepackage{graphicx}
\usepackage{makecell}
\usepackage{cleveref}
\usepackage{lscape}
\usepackage{pdflscape}
\usepackage{longtable}
\usepackage{rotating}
\usepackage{graphicx}
\usepackage{adjustbox}

\DeclareMathOperator*{\argmin}{arg\,min}

\newcommand{\ie}{{\it i.e., }}
\newcommand{\eg}{{\it e.g., }}
\newcommand{\etc}{{\it etc. }}

\title{\LARGE{Robust Portfolio Selection Problems: A Comprehensive Review}}
\author[1]{Alireza Ghahtarani\thanks{Dalhousie University, Halifax, Nova Scotia, B3J 1B6, Canada (alireza.ghahtarani@dal.ca).}}
\affil[1]{Department of Industrial Engineering, Dalhousie University}
\author[1]{Ahmed Saif}
\author[1]{Alireza Ghasemi}
\date{\vspace{-1.5cm}}
\begin{document}
\maketitle

\abstract{This paper reviews recent advances in robust portfolio selection problems and their extensions, from both operational research and financial perspectives. A multi-dimensional classification of the models and methods proposed in the literature is presented, based on the types of financial problems, uncertainty sets, robust optimization approaches, and mathematical formulations.  Several open questions and potential future research directions are identified.

}

\section{Introduction}

The portfolio selection problem (PSP) is a fundamental problem in finance that aims at optimally allocating funds among financial assets to maximize return and/or minimize risk. Different variants of the problem arise in reality due to the different risk attitudes of investors (risk-neutral \emph{vs.} risk-averse), investment strategies, measures used to quantify risk (\eg variance, VaR), methods used to calculate return (\eg log-return) and planning horizon (single-period \emph{vs.} multi-period), among other factors. Consequently, the PSP literature has grown considerably in terms of both size and diversity, allowing for several classification schemes to be employed. 
 
An obvious classification is based on the risk measure to be optimized. Generally speaking, two broad classes of risk measures have been proposed: volatility-based and quantile-based. While variance has been the most widely-used risk measure in both theory and practice since the seminal work of \cite{markowitz1952portfolio}, it has its deficiencies. First, it equally considers both positive and negative deviations around the expected return as undesirable risk, despite the desirability of the positive deviations for investors. Alternatively, downside risk measures that consider only the negative deviations of returns, like the lower partial moment (LPM), can be used. Furthermore, given that variance is a nonlinear risk measure, it leads to more complex formulations than those corresponding to linear risk measures like the mean absolute deviation (MAD) proposed by \cite{konno1991mean}. Related to volatility risk measures, \cite{sharpe1966mutual} and \cite{bernardo2000gain}, introduced \emph{Sharpe ratio} and \emph{Omega ratio}, respectively, to evaluate the performance of portfolios based on risk and return simultaneously. The most famous quantile-based risk measures are \emph{Value-at-Risk} (VaR) and \emph{Conditional-Value-at-Risk} (CVaR). The former quantifies the maximum loss at a specific confidence level, whereas the latter represents the expected value of losses greater than VaR at a confidence level. For details about quantile-based risk measures, interested readers are referred to \cite{rockafellar2000optimization}. 

Besides risk measures, PSPs can be classified based on investment strategies. For example, index tracking, first studied by \cite{dembo1992tracking}, is a passive investment strategy that tries to follow a market index. On the other hand, active investment strategies that involve ongoing buying and selling of assets are optimized by solving multi-period PSPs (see \cite{dantzig1993multi} for an early example). %The asset liability management (ALM) problem is a special case of the multi-period PSP which aims to ensure that liabilities in financial institutions are covered. 
Furthermore, hedging gives rise to a popular PSP in which an investment position is intended to offset potential losses or gains that may be incurred by a companion investment. Interested readers are referred to \cite{lutgens2006robust} for a detailed account of financial hedging strategies. PSPs can be classified also according to return calculation methods. \cite{goldfarb2003robust} incorporated factors (macroeconomic, fundamental, and statistical) to determine market equilibrium and calculate the required rate of return, whereas \cite{hull2003options} defined the \emph{Log-return} as the equivalent, continuously-compounded rate of return of asset returns over a period of time.

Despite being a well-studied problem, a common feature of most PSPs addressed in the literature is that the problem parameters are assumed to be known with certainty. Ignoring the inherent uncertainty in parameters and instead using their point estimates often leads to suboptimal solutions. Two widely-used frameworks for dealing with uncertainty are \emph{stochastic programming} (SP) and \emph{robust optimization} (RO). SP focuses on the long-run performance of the portfolio by finding a solution that optimizes the expected value of the loss function. Despite its intuitive appeal and favorable convergence properties, SP requires the distribution function of the uncertain parameters to be known. Moreover, its risk-neutral nature does not provide protection from unfavorable scenarios, rendering it unsuitable for, typically, risk-averse investors. On the other hand, RO is a conservative approach that minimizes the loss function under the worst-case scenario (within an \emph{uncertainty set}) and does not use information about the probability distribution of the uncertain parameters, making it an attractive alternative.

Given the rising interest in robust PSPs in the last two decades, several attempts have been made to review the growing robust PSP literature. Among the earliest reviews is that of \cite{fabozzi2010robust}, which concentrates on the application of RO on basic mean-variance, mean-VaR, and mean-CVaR problems, but does not cover more recent variants of the problem like robust index tracking, robust LPM, robust MAD, robust Omega ratio, and robust multi-objective PSPs. \cite{scutella2010robust} and \cite{scutella2013robust} also review robust mean-variance, robust VaR, and robust CVaR problems, but similarly, do not survey other robust PSPs. Likewise, \cite{kim2014recent} concentrate on worst-case formulations, while ignoring other important classes, including relative robust models, robust regularization, net-zero alpha adjustment and asymmetric uncertainty sets. Another review by \cite{kim2018recent} focuses on worst-case frameworks in bond portfolio construction, currency hedging, and option pricing, while covering a small number of references on robust asset-liability management problems, log-robust models, and robust multi-period problems.
Recently, \cite{xidonas2020robust} provided a categorized bibliographic review which broadly covers the area; their aim is to provide a rapid access to the topic for finance practitioners, and in general for those interested, but maybe not yet in the area.

\begin{figure}[H]
    \centering
    \includegraphics[width=18cm]{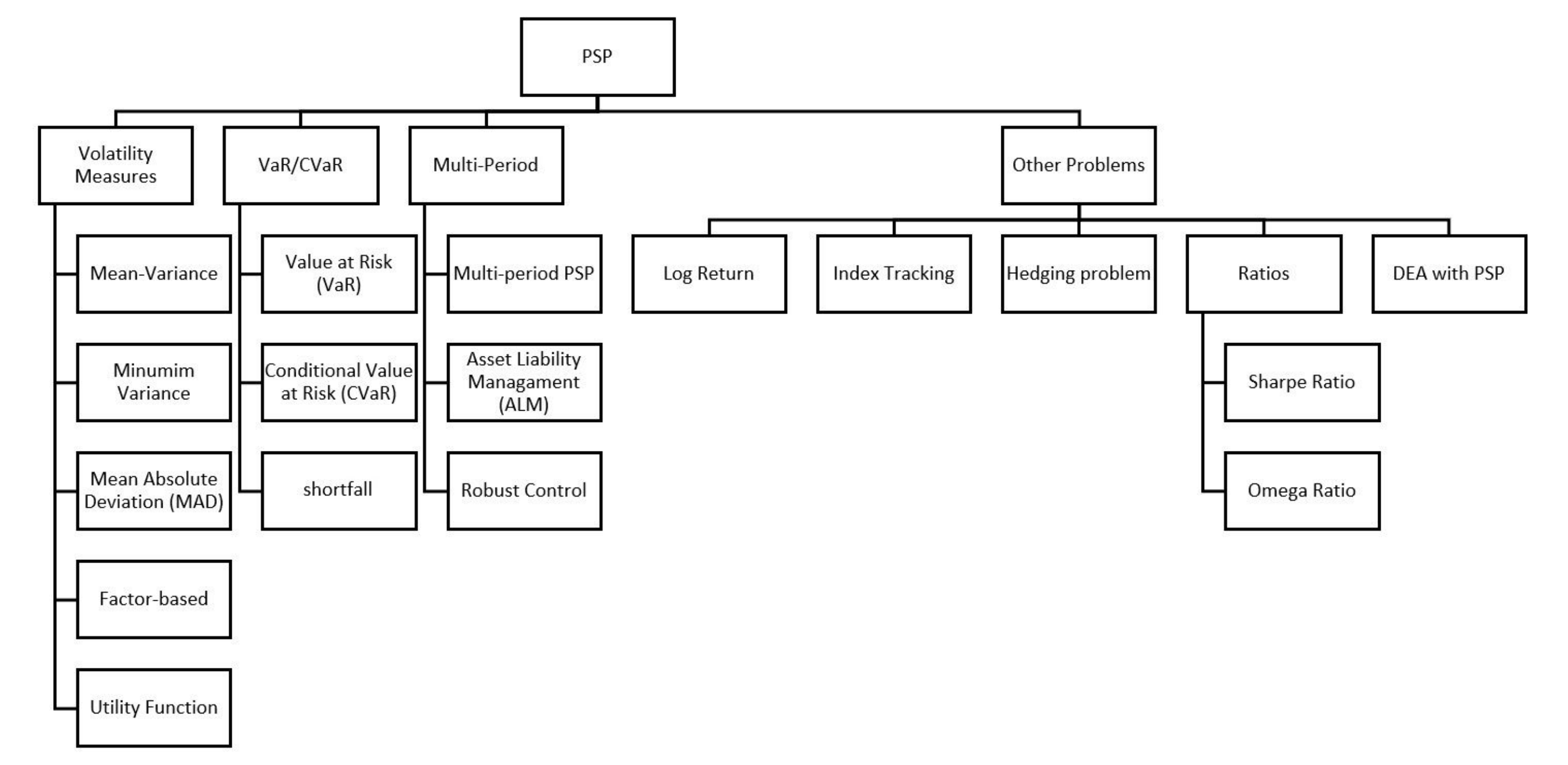}
     \caption{A schematic diagram of the classification scheme of robust PSPs utilized in this review}
     \label{fig:classification}
\end{figure}

The main contribution of this review paper is a multi-dimensional classification of robust PSPs. The classification scheme of robust PSPs utilized in this review is illustrated in Figure \ref{fig:classification}. To put together the list of references to be reviewed, we first compiled two sets of keywords. The first set includes the following keywords related to financial problems: “portfolio selection”, “risk measures”, “VaR”, “CVaR”, “mean-variance”, “semi-variance”, “mean absolute deviation”, “index tracking”, “factor-based portfolio”. The second set includes the robust optimization keywords: “robust optimization”, “distributionally robust optimization”, “data-driven”, and “uncertainty set”.
We then searched all pairs/combinations of the first and second keyword sets on both \emph{Scopus} and \emph{Web of Science databases}, and also using the \emph{Google Scholar} search engine. The references retrieved from these searches were carefully screened to make sure that they are related to the robust financial problems. If a paper was deemed related to the scope of this review, we searched the references cited in it and those that cited it to find additional references to be included. This process was repeated multiple times until no new references could be found.

\begin{figure}[H]
    \centering
    \includegraphics[width=14cm]{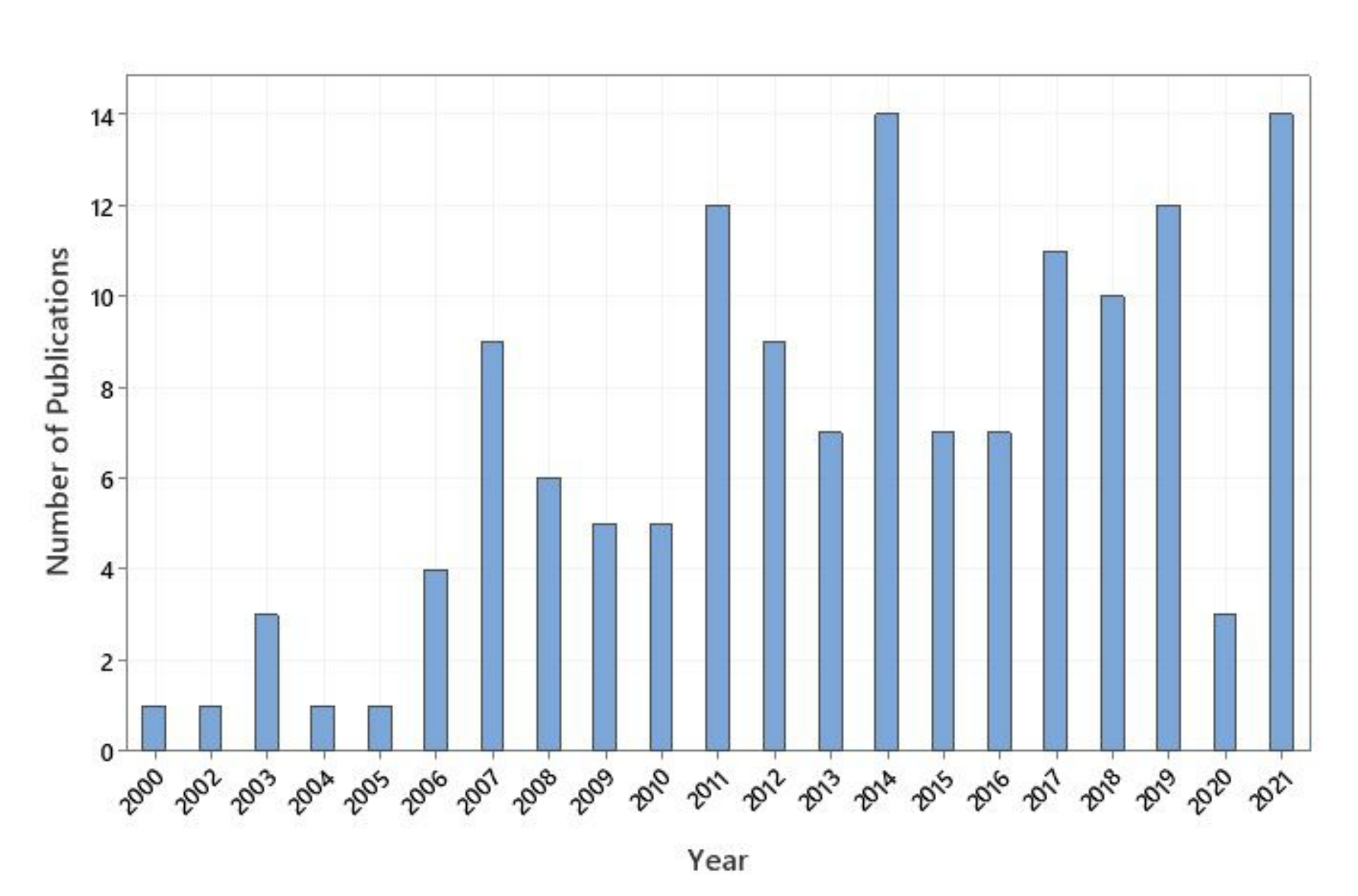}
     \caption{A breakdown of the reviewed article by publication year}
          \label{fig:by year}
\end{figure}

\begin{figure}[H]
    \centering
    \includegraphics[width=18cm]{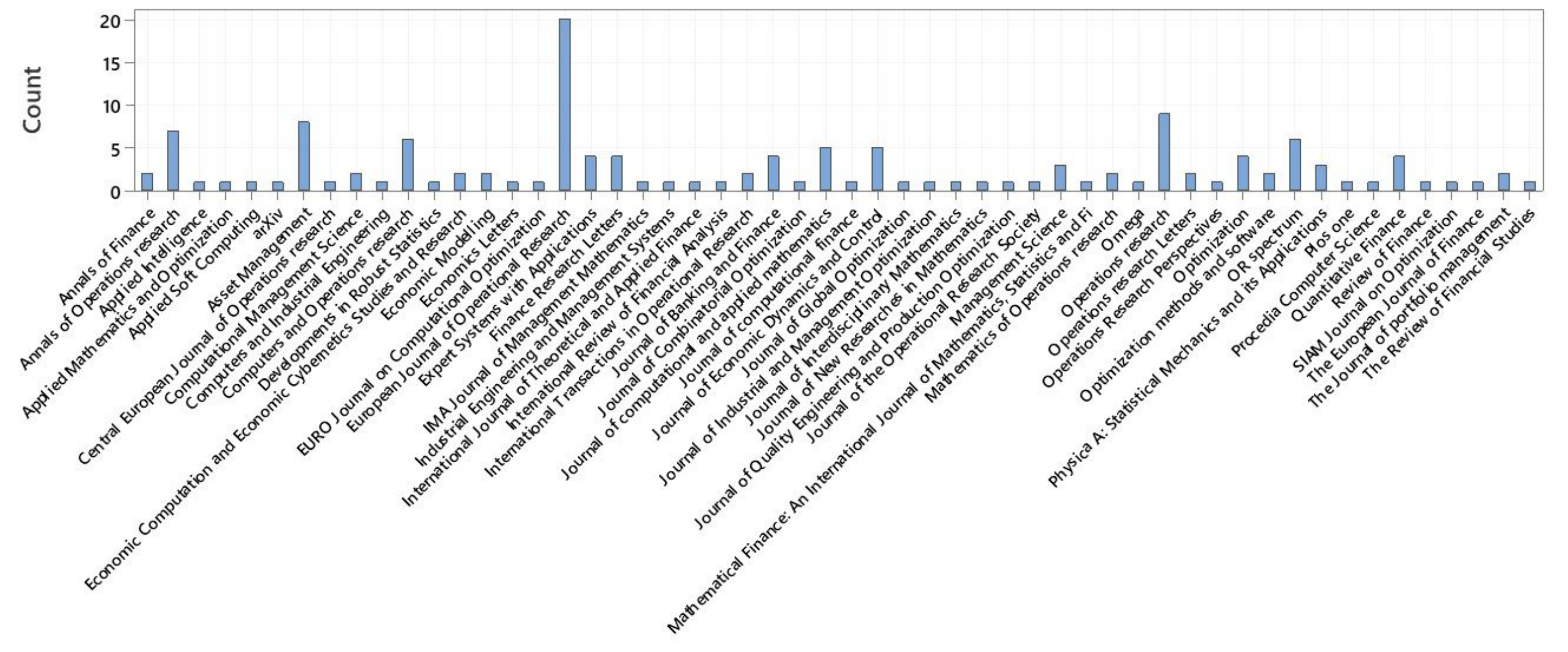}
     \caption{A breakdown of the reviewed articles by journal}
          \label{fig:by journal}
\end{figure}

Figure \ref{fig:by year} portrays a breakdown of the reviewed articles by publication year, spanning between the years 2000 and 2021. We note that out of the 142 articles reviewed, 14 appeared in 2021, thus were not included in any of the previous reviews. Our review focuses on articles published in peer-reviewed journals. These articles appeared in a large number ($n=54$) of finance and operations research (OR) journals. Figure \ref{fig:by journal} shows a breakdown of the reviewed papers by journal (sorted alphabetically). We note that most robust PSP articles were published in OR journals.

A major challenge when reviewing the robust PSP literature is the absence of a unified set of nomenclatures and notations for describing and formulating the problems. To be able to link and contrast different variants of the problem, we use, throughout our review, a unified set of most used notations, shown in Table \ref{table:1}. The notations that are used once are defined in the text. Our strategy for including mathematical formulas was to begin with the simplest and most general ones, then incrementally add new or alternative items (\eg terms in the objective function, constraints, risk measures, levels of optimization) at their first use in the robust PSPs literature. We also chose to include formulas that are commonly used and that constitute significant contributions, leaving behind some outliers and minor changes for brevity.

\begin{center}
\begin{longtable}{|c | l|} 
\caption{Notations and symbols}
\label{table:1}
 \\ \hline 
 Symbol/Notation & Definition\\ [0.5ex] 
 \hline\hline
\(x \in \mathbb{R}^{n}\)	&	Decision variable, \(x_{j}\) the proportion of the available budget invested in asset \(j\)	\\
\hline
\(r \in \mathbb{R}^{n}\)	&	Asset return vector	\\
\hline
\(r^{L} \in \mathbb{R}^{n}\)	&	Minimum asset return vector	\\
\hline
\(r^{U} \in \mathbb{R}^{n}\)	&	Maximum asset return vector	\\
\hline
\(r_{f} \in \mathbb{R}\)	&	Risk-free asset return	\\
\hline
\(Q\in \mathbb{R}^{n\times n}\)	&	Variance-covariance matrix of the assets	\\
\hline
\(Q^{L}\in \mathbb{R}^{n\times n}\)	&	Minimum variance-covariance matrix of the assets	\\
\hline
\(Q^{U}\in \mathbb{R}^{n\times n}\)	&	Maximum variance-covariance matrix of the assets	\\
\hline
\(E \in \mathbb{R}\)	&	Portfolio expected return	\\
\hline
\(v \in \mathbb{R}\)	&	Portfolio variance	\\
\hline
\(e\)	&	Vector of size \(n\) whose components are ones	\\
\hline
\(\lambda \in \mathbb{R}\)	&	Risk aversion coefficient	\\
\hline
\(U_{Q}\)	&	Uncertainty set of \(Q\), which has the same dimensions of the uncertain parameter \\
\hline
\(U_{r}\)	&	Uncertainty set of \(r\), which has the same dimensions of the uncertain parameter	\\
\hline
\(\Gamma \in \mathbb{R}^{+}\)	&	Non-negative scalar that controls the size of uncertainty set	\\
\hline
\(Pr_{j} \in \mathbb{R}\)	&	Price of asset \(j\)	\\
\hline
\(E_{i} \in \mathbb{R}\)	&	Exchange rate of currency \(i\)	\\
\hline
\(\Sigma \in \mathbb{R}^{n\times n}\)	&	Covariance matrix of the estimated expected returns	\\
\hline
\(\pi\)	&	Nominal distribution function	\\
\hline
\(p\)	&	True distribution function	\\
\hline
\(\zeta\)	&	A random variable	\\
\hline
\(A \in \mathbb{R}^{n \times n}\)	&	A positive semi-definit matrix	\\
\hline
\(\eta_{i}\)	&	A positive scalar	\\
\hline
\(t\)	&	Indices of scenarios	\\
\hline
\(k_{min} \in \mathbb{R}^{n}\)	&	Lower bound of decision variables	\\
\hline
\(k_{max} \in \mathbb{R}^{n}\)	&	Upper bound of decision variables	\\
\hline
\(c_{j}\)	&	Binary variable, if  the  asset \(j\) is  selected  it  takes  one,  otherwise zero	\\
\hline
\(L \in \mathbb{Z}\), \(H \in \mathbb{Z}\)	&	Integer scalars that show the minimum and maximum number of assets in the portfolio	\\
\hline
\(s\)	&	Indices of period	\\
\hline
\(W_{0} \in \mathbb{R}\)	&	Initial wealth of the investors	\\
\hline
\(\beta \in \mathbb{R}\)	&	Confidence level	\\
\hline
VaR	&	Value at Risk	\\
\hline
CVaR	&	Conditional Value at Risk	\\
\hline
\(f(x,r)\)	&	Loss function	\\
\hline
\(\Delta \in \mathbb{R}\), \(\nu \in \mathbb{R}\)	&	Transaction costs (buying and selling)	
\\ [1ex] 
 \hline

\end{longtable}
\end{center}

The remainder of this review paper is organized as follows. The next section provides a brief introduction to RO for non-specialists. Section \ref{volatility} surveys robust PSP formulations based on volatility measures. Section \ref{quantile} reviews quantile-based PSPs, which include Value at Risk (VaR), Conditional Value at Risk (CVaR), and their extensions with worst-case RO methods, relative RO and distributionally robust optimization (DRO). Furthermore, the relationship between uncertainty sets and risk measures, application of soft robust formulation with risk measures, worst-case CVaR and its relationship with uniform investment strategy, and robust arbitrage pricing theory with worst-case CVaR are also discussed in Section \ref{quantile}. Section \ref{multi-period} provides a review of RO in multi-period PSPs and asset-liability management (ALM) problems. Besides these two main problems, robust control formulations of investment problems are reviewed in this section. Section \ref{sec:special} reviews other financial problems that are not covered in the above-mentioned categories like robust log-return, index-tracking, hedging problem, risk-adjusted Sharpe ratio, robust scenario-based formulation, and robust data envelopment analysis. The last section provides conclusions and open issues in this context.     

\section{A Brief Introduction to Robust Optimization} \label{RO}

This section provides a brief introduction to RO for readers who are not familiar with the topic.  RO is a framework for dealing with the uncertainty of parameters in optimization problems by assuming that the parameters belong to an \emph{uncertainty set} and optimizing over the worst realization in this set. The first RO formulation was developed by \cite{soyster1973convex} and used a \emph{box} (hypercubic) uncertainty set that specifies an interval for each individual uncertain parameter. Even though this approach usually leads to tractable formulations, it is too conservative since it is based on the assumption that all parameters will take their worst possible values simultaneously, which rarely happens in reality. To overcome this issue, \cite{ben1998robust} proposed an \emph{ellipsoidal} uncertainty set that is centered at some nominal value and has a size (radius) that controls the conservatism of the solution based on the decision maker's aversion to uncertainty. Nevertheless, tractable reformulations of RO problems with ellipsoidal uncertainty sets give rise to nonlinear formulations that, generally, have a higher complexity than the nominal problem. Later, \cite{bertsimas2004price} developed a special class of polyhedral uncertainty set, referred to as \emph{budget}, that enables the level of conservatism to be controlled while preserving the tractability of the reformulated problems. All of the aforementioned uncertainty sets are \emph{symmetric}, meaning that they are based on the assumption that forward and backward deviations around the nominal value are equal. \cite{chen2007robust} argued that this assumption is not valid in many practical settings and proposed an \emph{asymmetric} uncertainty set that is particularly suitable for financial applications. 

Despite the protection it provides against adverse scenarios, classical RO is still considered overly conservative and pessimistic by many practitioners. To alleviate this concern, several RO variants have been developed. \cite{scherer2007can} proposed adding a net-zero alpha adjustment constraint to any uncertainty set to guarantee that for any downward adjustment in the uncertain parameter, there is an offsetting upward adjustment, thus reducing the level of conservatism. \cite{kouvelis1997robust} proposed a \emph{relative} RO approach that uses a regret function under the least desirable scenario. Although this approach provides solutions that perform better, on average, than classical RO, it suffers from intractability since it results in a three-level optimization problem. 

Another way to achieve less conservative solutions is to use available partial information about the distribution function of the uncertain parameters rather than completely overlook them. A framework referred to as \emph{distributionally robust optimization} (DRO), that dates back to the seminal work of \cite{scarf1958min}, has gained considerable attention recently. It assumes that the unknown probability distribution of the uncertain parameters belongs to a set of distributions called the \emph{ambiguity set}, and optimizes the expected value of the objective function, where the expectation is taken with respect to the worst distribution in this set. Clearly, the tractability, convergence and out-of-sample performance guarantee offered by the DRO solution obtained depends on the ambiguity set used. Generally speaking, there are two main types of ambiguity sets: moment-based and discrepancy-based. The former includes distributions that enjoy some parametric properties, \eg mean or variance, whereas the latter include distributions that are within a certain ``distance" (\eg the \emph{Kullback–Leibler divergence} or the \emph{Wasserstein metric}) from a reference distribution. The interested reader is referred to \cite{rahimian2019distributionally, esfahani2018data, delage2010distributionally} and the references therein for more information about DRO.

\section{Robust PSPs with Volatility-based Risk Measures}\label{volatility}

In this section, we review the application of RO in PSPs with volatility-based risk measures, which include mean-variance, mean absolute deviation, lower partial moment, systematic risk, Omega ratio, and factor-based portfolio models.

\subsection{Mean-Variance PSP}

The mean-variance PSP was proposed by \cite{markowitz1952portfolio}. In its general form, it assumes \(n\) risky assets, each has an expected rate of return denoted by the vector \(r\), whereas \(v\) is the portfolio variance and \(Q\) is the variance-covariance matrix of the assets. In Markowitz's model, the variance of the portfolio is the risk measure to be optimized. The decision variable of this mathematical formulation is \([x_{j}]_{j=1,\dots,n}\), which represents the proportion of the available budget invested in asset \(j\). When \(x_{j}\geq 0\), it means that short selling is not allowed. 

Moreover, \(E=x'r\) is the portfolio expected return, \(v=x'Qx\) is the portfolio variance, and \(E_{0}\) is the minimum required expected rate of return. Then, the minimum variance PSP is \(\min\limits_{x\in X} (v=x'Qx)\;\text{s.t.}\;x'r\geq E_{0}\) where \(X=\{x:\;e'x=1,\;x_{j}\geq 0,\;\;j=1,...,n\}\), and \(e\) is a vector of size \(n\) whose components are ones. 
Another version of the mean-variance PSP, called the risk-adjusted expected return, takes the form \(\max\limits_{x\in X} (x'r-\lambda x'Qx)\). This formulation has the dual objectives of maximizing the portfolio return and minimizing its variance, where \(\lambda\) is a risk aversion coefficient set by the investor. In reality, however, the true values of the expected rate of return and the covariance matrix are not known with certainty.

\subsubsection{Classical Uncertainty Sets for the Mean-Variance PSP}

The general robust counterparts of the aforementioned PSPs are \(\min\limits_{x \in X} \max\limits_{Q\in U_{Q}}(x'Qx)\;\text{s.t.}\min\limits_{r\in U_{r}}x'r\geq E_{0}\), 
and \(\max\limits_{x \in X}\min\limits_{r\in U_{r},\; Q\in U_{Q}} (x'r-\lambda x'Qx)\), respectively, where \(U_{Q}\) and \(U_{r}\) are the uncertainty sets for the covariance matrix and the return vector, respectively. \cite{tutuncu2004robust} used symmetric box uncertainty sets defined as \(U_{r}:=\{r:\; r^{L}\leq r\leq r^{U}\}\) and \(U_{Q}=:\{Q:\; Q^{L}\leq Q \leq Q^{U}, Q\succeq 0\}\),
where \(r^{L}\) and \(r^{U}\) are, respectively, the lower and upper bounds of the asset returns and \(Q^{L}\) and \(Q^{U}\) are the lower and the upper bounds of the covariance matrix elements while stipulating also that \(Q\) must remain positive semi-definite (PSD). It has been shown that, with these uncertainty sets, the robust counterparts could be tractably formulated as \(\min\limits_{x\in X}(x'Q^{U}x)\;\text{s.t.}\;x'r^{L}\geq E_{0}\), and \(\max\limits_{x \in X}(x'r^{L}-\lambda x'Q^{U}x)\), respectively. \cite{khodamoradi2020robust} used similar uncertainty sets for a cardinality-constrained mean-variance PSP that allows short selling. \cite{swain2021robust} also analyzed the robust mean-variance, and mean-semi-variance PSPs with box uncertainty sets, where both the expected return vector and the covariance matrix are uncertain parameters. However, \cite{chen2009robust} argued that deviations of the expected asset returns from their nominal values are not symmetric, meaning that the upside deviation is different from the downside deviation, thus are not accurately captured by classical symmetric uncertainty sets. Instead, they used non-symmetric interval uncertainty sets for the expected vector and covariance matrix of asset returns. The element-wise uncertainty interval was defined as \(U_{r_{i}}=[\bar{r}_{i}-\theta_{i}^{1},\bar{r}_{i}+\theta_{i}^{2}]\) and \(U_{q_{ij}}=[\bar{q}_{ij}-\tau_{ij}^{1},\bar{q}_{ij}+\tau_{ij}^{2}]\),
where \(\bar{r}_{i}\) and \(\bar{q}_{ij}\) are elements of \(r\) and \(Q\), respectively, that represent the nominal values of mean and covariance, whereas  \(\theta_{i}^{1}\) and \(\theta_{i}^{2}\) are the downside and the upside deviations for the mean and  \(\tau_{ij}^{1}\) and \(\tau_{ij}^{2}\) are the downside and the upside deviations for the covariance, respectively. To propose a robust counterpart, optimistic \(f_{opt}\) and pessimistic \(g_{pes}\) values are defined as \(f_{opt}=\min_{r_{i}\in U_{r_{i}}}(x'r)\), and \(g_{pes}=\max_{q_{ij}\in U_{q_{ij}}}(x'Qx)\), respectively. Alternatively, \cite{fabozzi2007robust} defined an ellipsoidal uncertainty set for asset returns as \(U_{r}:=\{r:\;(r-\bar{r})Q^{-1}(r-\bar{r})'\leq \Gamma^{2}\}\), where $\hat{r}$ is the nominal return and \(\Gamma^{2}\) is a non-negative scalar that controls the size of the uncertainty set. Hence, the robust counterpart can be tractably formulated as \(\{\min\limits_{x\in X}{-\bar{r}'x+\Gamma \sqrt{x'Qx}+\lambda x'Qx}\}\). However, the uncertainty of the covariance matrix was not considered, making the solution robust only against perturbations in the return vector. \cite{pinar2016robust} developed a robust mean-variance PSP with the same ellipsoidal uncertainty set while allowing short selling, which was also extended to the multi-period case. 

Even though RO accounts for uncertainty in the problem parameters, 
\cite{zymler2011robust} argued that if the uncertainty set is not set large enough, the solution might maintain its robustness only under normal market conditions, but not when the market crashes. Instead, they proposed using European-style options to hedge the mean-variance portfolios against abnormal market conditions. Two guarantee types were provided: weak, and strong. The weak guarantee applies under normal market conditions when the rate of the return is varying in an ellipsoid uncertainty set, whereas the strong guarantee applies to all possible asset returns by using the European-style options in the form of constraints in the optimization problem. Hence, the strong guarantee is not based on RO formulation but on the mechanism of options to protect the portfolio in market crashes. \cite{ashrafi2021study} also used the idea of strong and weak guarantees. For the strong guarantee, an option is used in PSP, whereas for the weak guarantee, a budget uncertainty set for asset returns is used. Hence, the problem can be reformulation as a linear program (LP).

According to \cite{lu2019asset}, an important drawback of the mean-variance PSP is that its inputs are computed using only historical market returns, thus specific earnings announcements cannot be used to support the portfolio selection process. To overcome this issue, \cite{black1990asset} proposed the BL method, which consists of both a market model and a view model. \cite{lu2019asset} improved the view model of the BL method by using fuzzy logic to make it quantitative. Moreover, they incorporated multiple expert views instead of just one individual expert view in their formulation. To handle the heterogeneity of data collected from disparate sources, they applied RO with an ellipsoidal uncertainty set for the mean return vector and the return covariance matrix.

\cite{fonseca2012robust} asserted that an important strategy in the PSP is diversification, which may eliminate some degree of risk since financial assets are less than perfectly correlated. To make a portfolio more diversified, investors can invest in foreign assets. However, foreign exchange rates' fluctuations may erode the investment's return. Moreover, both the asset returns and the currency rates are uncertain. Hence, \cite{fonseca2012robust} and \cite{fonseca2012robusta} proposed a robust formulation for the international PSP with an ellipsoidal uncertainty set, which leads to a non-convex bilinear optimization problem. The problem considers \(n\) assets from \(m\) foreign currencies, where \(Pr_{j}^{0}\) and \(Pr_{j}\) are, respectively, the current and future prices of asset \(j\), and \(E_{i}^{0}\) and \(E_{i}\), respectively, are the current and future exchange rates of currency \(i\). Therefore, the local return of asset $j$ is \(r_{j}^{a}=Pr_{j}/Pr_{j}^{0}\) and the exchange rate return of currency $i$ is \(r_{i}^{e}=E_{i}/E_{i}^{0}\). Using the auxiliary binary matrix \(O=[o_{ji}]\), where \(o_{ji}\) equals 1 if asset $j$ is traded in currency $i$ and 0 otherwise, the international PSP is formulated as \(\max_{x \in X} \min_{(r^{a},r^{e})\in U_{r^{a},r^{e}}}{[diag(r^{a}Or^{e})]'x}\), where the objective of this formulation is to maximize the worst-case return within all realizations in the uncertainty set \(U_{r^{a},r^{e}}\). A semi-definite programming (SDP) approximation is proposed to handle the non-linearity of the robust international PSP. Even though a robust international portfolio provides some level of guarantee against the uncertainty, investors might alternatively use forward contracts and quanto options (an exotic type of options translated at a fixed rate into another currency) to hedge risk. To make the formulation more practical, \cite{fonseca2012robust} extended the robust international PSP with forward contracts and quanto options to reach a less conservative formulation. 

Another classical uncertainty set used for robust PSPs is the budget uncertainty set proposed by \cite{bertsimas2004price}, which has the advantage of leading to tractable reformulations with the same complexity of the nominal problems. \cite{sadjadi2012robust} considered a robust cardinality constrained mean-variance PSP with ellipsoidal, budget, and general-norm uncertainty sets and proposed a genetic algorithm to solve them. It was shown that using a budget uncertainty set has led to better rates of return compared to other uncertainty set types. \cite{gregory2011robust} also tested a budget uncertainty set for the uncertain returns in a PSP to show the impact of the uncertainty set size on the portfolio return. The uncertainty set in this formulation is defined as \(U_{r}=\{r:\;r=\bar{r}+\hat{r}\zeta,\; ||\zeta||_{1}\leq \Gamma,\; |\zeta|\leq 1 \}\), where \(\bar{r}\) is nominal value, \(\hat{r}\) is the deviation of return, \(\zeta\) is the random variable, and \(\Gamma\) is the price of robustness that control the size of uncertainty set. The final formulation is  \(\max_{x\in X,\;z\geq 0,\:q\geq 0}{\bar{r}'x-\Gamma z-e'q}\;s.t.\; z+q_{i}\geq \hat{r}_{i}x_{i},\;\forall i\). It has been shown that using the mean or the median of the asset returns as nominal values leads to the most robust portfolios.

\cite{bienstock2007histogram} postulates that the solution methodology of RO is often chosen at the expense of the accuracy of the uncertainty model. Moreover, classical uncertainty sets might lead to overly conservative solutions. Alternatively, \cite{bienstock2007histogram} proposed a data-driven approach to construct the uncertainty set by using uncertainty bands, each showing a different level of the return shortfall, which is an amount by which a financial obligation or liability exceeds the required amount of cash that is available. Hence, it is possible to specify rough frequencies of return shortfalls to approximate the return shortfall distribution. The robust models are formulated by allowing the uncertain parameter (asset returns) to deviate from the distribution by incorporating constraints related to the frequency of the return shortfall in different bands, an approach referred to as \emph{robust histogram mean-variance} PSP. Although this formulation provides more flexibility than classical RO, the robust counterpart is an intractable mixed-integer program (MIP); thus, a cutting plane algorithm is proposed to solve it.

\subsubsection{Robust Mean-Variance and Multi-Objective Solution Methods}

\cite{fliege2014robust} studied a robust multi-objective optimization (MOO) version of the mean-variance PSP while considering minimizing the variance and maximizing the mean return of the portfolio as the two objectives. Two MOO methods were applied: the \(\varepsilon\)-constraint scalarization (ECS) method, which pushes one of the objective functions, namely return maximization, to the constraints, and the weighted-sum scalarization (WSS) method, which combines the two objectives into a single one by assigning proper weights to them. Both methods lead to the same efficient frontier in the nominal case, but not for the robust problem. \cite{fliege2014robust} defined the location characteristics of the robust Pareto frontier with respect to the non-robust Pareto frontier, and demonstrated that standard techniques (ECS and WSS) from MOO can be used to construct the robust efficient frontier.

Alternatively, robustness in multi-objective PSPs can be achieved by a resampling method without classical uncertainty sets, which provides a wider range for uncertain parameters and solutions instead of the worst-case scenario RO with a specific uncertainty set. 
This approach requires replacing the parameters in the fitness functions at every generation. Hence, the evolution process would favor the solutions that show good performance in terms of risk and return over different scenarios (see \eg \cite{shiraishi2008resampling}, \cite{ruppert2014statistics}). \cite{garcia2012time} argued that one of the main problems portfolio managers face is uncertainty regarding the expected frontier derived from forecasts of future returns. Very often, expected frontiers lie far from the actual return, resulting in inaccurate forecasts of the portfolio risk/return profile. 
\cite{garcia2012time} demonstrated that robustness of results can be achieved by avoiding optimization based on a single expected scenario that may produce solutions that are hyper-specialized and might be extremely sensitive to likely deviations. They tackle the problem of achieving robust or stable portfolios by using a multi-objective evolutionary algorithm that replaces the traditional fitness function with an extended one that uses a resampling mechanism and an implicit third objective to control the robustness of the solutions.

The formulations of \cite{fliege2014robust} and \cite{garcia2012time} are based on differentiable functions. However, classical RO methods cannot be used on nonsmooth and non-differentiable functions. To address this issue, \cite{fakhar2018nonsmooth} developed the necessary and sufficient optimality condition for a MOO problem with nonsmooth, \eg non-differentiable or discontinuous, functions, and proved that strong duality holds when these functions are convex.  Moreover, they introduced the concept of saddle-point for MOO under uncertainty. 

\subsubsection{Robust Optimization Based on the Estimation Error}

\cite{ceria2006incorporating} demonstrated that the mean-variance PSP is very sensitive to small variations in expected returns. They, instead, proposed a formulation for the robust PSP based on estimation errors. In this formulation, three distinctive Markowitz efficient frontiers were introduced: the true frontier calculated based on the true, yet unobservable, expected returns, the estimated frontier calculated based on the estimated return, and the actual frontier calculated based on the true expected returns of the portfolios on the estimated frontier. To have a portfolio as close as possible to its true frontier, the maximum difference between the estimated frontier and the actual frontier was minimized. \cite{ceria2006incorporating} modified the maximum difference between the estimated frontier and the actual frontier by adding a linear constraint. 
They assumed that the true returns lie inside the confidence region \((r-\bar{r})'\Sigma^{-1}(r-\bar{r})\leq k^{2}\), where \(k^{2}\sim \chi_{n}^{2}\), and \(\chi_{n}^{2}\) is the inverse cumulative distribution function of the chi-squared distribution with \(n\) degrees of freedom. Points on the efficient frontier are calculated by solving \(\max{\bar{r}'x}\;\text{s.t.}\;x'Qx\leq v\), where \(v\) is the maximum acceptable variance. The optimal solution of this optimization problem is \(x=\sqrt{\frac{\nu}{r'Q^{-1}r}}Q^{-1}r\). 
By considering \(r^{*}\) as the true, but unknown, expected return vector and \(\bar{r}\) as an expected return, the true expected return of a portfolio on the estimated frontier is computed as \(\sqrt{\frac{\nu}{\bar{r}{'}Q^{-1}\bar{r}}}r^{*}{'}Q^{-1}\bar{r}\). \cite{ceria2006incorporating} assumed that \(\tilde{x}\) is the optimal portfolio on the estimated frontier for a given target risk level. Then, the difference between the estimated expected return and the actual expected return of \(\tilde{x}\) is \(\bar{r}{'}\tilde{x}-r^{*}{'}\tilde{x}\). Consequently, the maximum difference between the expected returns on the estimated efficient frontier and the actual efficient frontier is computed by solving \(\max\;\ (\bar{r}'\tilde{x}-r'\tilde{x})\;\text{s.t.}\;(r-\bar{r})'\Sigma^{-1}(r-\bar{r})\leq k^{2}\).
In this formulation \(\tilde{x}\) is fixed and optimization is over \(r\). Hence, the optimal solution is \(r=\bar{r}-\sqrt{\frac{k^{2}}{\tilde{x}^{'}\Sigma\tilde{x}}}\Sigma\tilde{x}\). Moreover, the lowest value of the actual expected return is \(\bar{r}'\tilde{x}=\bar{r}'\tilde{x}-k||\Sigma^{1/2}\tilde{x}||\). Finally, the maximum difference between the estimated frontier and the actual frontier is \(\bar{r}'\tilde{x}-(\bar{r}'\tilde{x}-k||\Sigma^{1/2}\tilde{x}||)=k||\Sigma^{1/2}\tilde{x}||\).
Effectively, minimizing the maximum difference between the actual and the true frontiers leads to a robust mean-variance PSP with an ellipsoidal uncertainty set, while the covariance matrix of estimation error is also captured using an ellipsoidal uncertainty set. 
\cite{garlappi2007portfolio} also claimed that the estimation error is ignored in mean-variance PSPs. They propose a robust mean-variance PSP that is a special case of the PSP in \cite{ceria2006incorporating} since asset returns are assumed to be normally distributed. Multiple historical data sets are used to estimate the random variable of asset returns. The problem is formulated as \(\max_{x\in X}\min_{r}{x'r-\lambda x'Qx}\;\text{s.t.}\;f(r,\bar{r},Q)\leq \varepsilon\), where \(\bar{r}\) is the estimated return, \(f(.)\) is a vector-valued function, \(\varepsilon\) is a vector of constants that captures the investor's uncertainty- and ambiguity-aversion. The additional constraint, representing the confidence interval of the normal assets return, shows that the decision maker accepts the possibility of estimation error. \cite{garlappi2007portfolio} compared their results for different \(f(.)\) selections with the results of the traditional Bayesian models and showed that their models are risk-averse while the Bayesian models are risk-neutral towards the uncertainty in parameters.

\subsubsection{Net-Zero Alpha Adjustment for the Mean-Variance}

\cite{scherer2007can} analyzed the results and models of robust estimation error by \cite{ceria2006incorporating} and robust mean-variance with box uncertainty set by \cite{tutuncu2004robust} and showed that the results of the robust mean-variance PSP are equivalent to the results of the mean-variance PSP with Bayesian shrinkage estimators for the uncertain parameters (for more details about Bayesian shrinkage estimators , see \eg \cite{lemmer1981ordinary}). The RO framework is criticized because it merely increases the complexity of the PSP while the solutions of the robust optimization, which depends on the choice of uncertainty set, are usually over-conservative. A method, referred to as \emph{net-zero alpha adjustment}, is developed, by which adding a constraint to the uncertainty set ensures that for any downward adjustment in the uncertain vector, there is an offsetting upward adjustment. For example, with the uncertainty set \(U=\{r=\bar{r}+\zeta: \zeta'\Sigma \eta\leq 1\}\), where \(\Sigma\) is the covariance matrix of estimation errors and \(\zeta\) is a deviation vector, the constraint \(e'\zeta=0\) is added. \cite{gulpinar2011robust} applied this method for a cardinality-constrained mean-variance PSP and found that adding a net-zero alpha adjustment to the ellipsoidal uncertainty set led to less conservative solutions than traditional robust mean-variance PSPs. 

\subsubsection{Distributionally Robust Mean-Variance} 

RO is a worst-case approach which assumes that the distribution function of uncertain parameters is unknown. However, partial information about the distribution function is often available, enabling less conservative distributionally robust optimization (DRO) formulations to be used. Several DRO models have been proposed for the mean-variance PSP. 

\cite{calafiore2007ambiguous} developed distributionally robust PSPs where two types of problems with different risk measures were addressed: the mean-variance PSP, which uses the mean and variance, and the mean absolute deviation PSP, which replaces the variance with the absolute deviation. 
Let us assume \(r(1),...,r(T)\) are \(T\) possible scenarios for the outcome of random return vector \(r\), and \(p_{t}\) is the probability associated to the scenario \(r(t)\), where \(\{p_{t}\geq 0,\; t=1,...,T,\;\sum_{t=1}^{T}p_{t}=1\}\).
Then, the expected value is defined as \(\mu(x,p)=\mathbb{E}[r'x]=\sum_{t=1}^{T}p_{t}r'(t)x=(\sum_{t=1}^{T}p_{t}r'(t))x=\bar{r}'(p)x\), where \(\bar{r}(p)=\mathbb{E}[r]=\sum_{t=1}^{T}p_{t}r(t)\). A risk measure can be quantified as the variance: \(v(x,p)=\mathbb{E}[(r'x-\mathbb{E}[r'x])^{2}]=x'Q(p)x\), where \(Q(p)\) is the covariance matrix of \(r\), and
\(Q(p)=\mathbb{E}[(r-\bar{r}(p))(r-\bar{r}(p))']=\sum_{t=1}^{T}p_{t}(r(t)-\bar{r}(p))(r(t)-\bar{r}(p))'\). 
Another risk measure in this concept is the expected absolute deviation (EAD) \(EDA(x,p)=\mathbb{E}[|r'x-E\{r'x\}|]=\sum_{t=1}^{T}p_{t}|r'(t)x-\mu(x,p)|\). By defining \(\lambda\geq 0\) as a risk aversion ratio, then the mean-variance PSP is \(\min_{x\in X} v(x,p)-\lambda \mu(x,p)\), and
the PSP based on the absolute deviation measure is \(\min_{x\in X}{EDA(x,p)-\lambda \mu(x,p)}\). A discrepancy-based ambiguity set based on the well-known Kullback-Leibler (KL) divergence, which measures the ``distance" between a nominal distribution vector \((\pi)\) and the unknown ``true" distribution vector \((p)\) is used, defined as \(KL(p,\pi)=\sum_{t=1}^{T}p_{t}log\frac{p_{t}}{\pi_{t}}\). Then \(p\) is only known to lie within KL distance \(d\geq 0\) from \(\pi\), \(K(\pi,d)=\{p:\;KL(p,\pi)\leq d\}\), where \(K(\pi,d)\) is the ambiguity set for the return distribution. This ambiguity set leads to a SDP formulation for the mean-variance PSP that is solvable using interior-point methods. The distributionally robust absolute deviation PSP is convex in the decision variable for any given distribution function. Consequently, a sub-gradient method combined with a proposed cutting plane scheme was used to solve the worst-case mean absolute deviation PSP in polynomial-time. \cite{baviera2021model} also applied KL divergence in the mean-variance PSP. However, unlike \cite{calafiore2007ambiguous}, they considered continuous distribution functions for the asset returns.

A limitation of \cite{calafiore2007ambiguous} is that, while the probabilities of scenarios are uncertain, the scenarios themselves are assumed to be known with certainty, which is not always the case in reality. \cite{pinar2014mean} formulated a semi-deviation PSP while considering uncertainty in both asset returns (through an ellipsoidal uncertainty sets) and in the distribution function of returns (through a moment-based ambiguity set). Both single and multi-period cases were considered.

As \cite{ding2018robust} argued, the Kullback–Leibler (KL) divergence used in \cite{calafiore2007ambiguous} is a special case of Rényi divergence with order one, hence they used it in a more general DRO formulation of the mean-variance PSP. Besides the risky assets having a multivariate normal distribution, they considered a risk-free asset with a fixed rate of return \(r_{f}\) in their formulation. By allowing short selling and using \(E\) as a target average return of the portfolio, the problem is formulated as \(\min_{x\in X}{x'Qx}\;\text{s.t.}\;x'(r-r_{f})\geq E-r_{f}\).
In the nominal case, the empirical distribution, obtained from historical data, is used , assuming that \(p\sim N(r,Q)\).
Since there is ambiguity about the true distribution of returns, an ambiguity set is constructed that contains all distributions within a certain distance, measured using Rényi divergence, from the empirical distribution. Renyi divergence is defined as \(D_{r}(p,\pi)=\frac{1}{r(r-1)}\ln{\int{p^{r}(\xi)\pi^{r-1}(\xi)d\xi}},\; r\neq 0,1\), 
where \(\pi(\xi)\) and \(p(\xi)\) are the probability density function under measures \(\pi\) and \(p\), respectively. Hence, the final formulation of the distributionally robust PSP is \(\max_{x\in X} \min_{r,Q}\; x'r-\lambda x'Qx+\lambda' D_{r}(p,\pi)\).
Their model was solved in three cases: only the mean return vector is uncertain, only the covariance matrix is uncertain, and both are uncertain.
It is worth mentioning that even though the ambiguity set used in \cite{ding2018robust} is more general than the KL-divergence, their formulations are special cases from the distribution function perspective since both the empirical and the true distribution function of the asset returns are assumed to be multivariate normal.

\subsubsection{Relative Robust Mean-Variance}

\cite{hauser2013relative} suggested that some professionals such as investment managers are frequently evaluated against their competitors and not on the absolute terms (worst-case solutions). The relative RO is the best possible approach to handle this situation where a regret function (the distance to the ``winner" under the least desirable scenario) is used to propose an intractable three-level optimization problem. \cite{hauser2013relative} incorporated a relative robust formulation into the mean-variance PSP where the regret function is \(Rgrt_{U,B}(x)=\max_{Q\in U} l_{B}(x,Q)=\max_{Q \in U}(\sqrt{x'Qx}-\min_{b\in B} \sqrt{b'Qb})\), \(x\) is the decision variable, \(Q\) is the variance-covariance matrix, \(U\) is an uncertainty set, and \(B={b_{1},...,b_{m}}\subseteq R^{n}\) is the set of benchmarks. To solve the proposed model, a polynomial-time solvable approximation for the inner problem was developed. 
The formulation of \cite{hauser2013relative} does not provide any control over regret value since the objective function is a regret function. Hence, \cite{article} extended a relative robust mean-variance PSP when a regret function is a constraint that provides more control over the regret value. Moreover, they defined proportional regret as an objective function, which is more perceivable by investors. Results show that the regret minimization seems to provide a greater degree of protection when it is compared to absolute robust optimization. \cite{caccador2021portfolio} proposed a new methodology for computing relative-robust solutions for mean-variance and minimum variance PSPs. This solution methodology is based on a genetic algorithm (GA), allowing the transformation of the three-level optimization problem into a bi-level problem.

\subsubsection{Robust Minimum Variance}

In the mean-variance PSP, it is assumed that there is a positive correlation between the expected return and the variance, which means more/less risk (variance) results in more/less profit (return). 
However, \cite{article2} showed that in a long-term investment strategy, low-volatility portfolios outperform high-volatility portfolios. Consequently, a PSP that minimizes just the variance of the portfolio (\ie global minimum variance portfolio) might have a better performance than the mean-variance PSP. \cite{MAILLET2015289} showed that the optimal solution of the global minimum variance portfolio can be calculated by solving a least-square regression while the covariance matrix of assets is uncertain. Hence, a robust least-square regression is proposed where the uncertainty set is based on the Frobenius norm, leading to a second-order cone program (SOCP).
\cite{MAILLET2015289} formulated the nominal global minimum variance PSP as \(\min_{x \in X}{x'\bar{Q}_Sx}\), where \(\bar{Q}_S\) is an estimate of the covariance matrix, leading to the closed-form optimal solution \(x^{*}=\frac{\bar{Q}_{S}^{-1}e}{e'\bar{Q}_{S}^{-1}e}\). They also showed that the optimal solution of this PSP can be computed as \(x^{*}=\frac{e}{n}-M \bar{\zeta}^{*}\),  where \(n\) is the number of stocks, \(M\) is an \(n\times (n-1)\) matrix having the following properties: \(M'e=0\) and \(M'M=I_{n-1}\), where \(I_{n-1}\) is the (\(n-1\)) identity matrix. \(\bar{\zeta}^{*}\) can be calculated based on the least square regression formulation \(\bar{\zeta}^{*}=\argmin_{\zeta}{||y-X\zeta||_{2}}\), where \(X=\bar{Q}_{S}^{1/2}M\), and \(y=\bar{Q}_{S}^{1/2}\frac{e}{n}\). Moreover, \(\bar{Q}_{S}^{1/2}\) is calculated from  \(\bar{Q}_{S}=\bar{Q}_{S}^{1/2}\bar{Q}_{S}^{1/2}\). For the robust PSP,
 \cite{MAILLET2015289} assumed that the pair \((X,y)\) is uncertain and belongs to a family of matrices \((X+\Delta X, y+\Delta y)\), where \(\Delta=[\Delta X, \Delta y]\) is a perturbation matrix while \(||\Delta||_{F}=||[\Delta X\; \Delta y]||_{F}\leq \rho\) is the uncertainty set, \(||.||_{F}\) is the Frobenius norm and \(\rho \geq 0\). Consequently, the robust counterpart of the least square regression is \(\bar{\zeta}(\rho)=\argmin_{\zeta} \max_{||\Delta X, \Delta y||_{F}\leq \rho}{||(y+\Delta y)-(X+\Delta X)\zeta||_{2}}\). Monte Carlo simulation was used to test the robust formulation, showing that it dominates the non-robust one with respect to weight stability, portfolio variance, and risk-adjusted returns. To make the formulation of \cite{MAILLET2015289} more practical, \cite{XIDONAS201760} augmented it into the cardinality-constrained global minimum variance PSP using the approach proposed by \cite{cornuejols2006optimization}, which uses scenarios instead of uncertainty sets to capture parameter uncertainty, making the formulation easier to handle. The problem is formulated as \(\min_{x\in X}x'Qx\;\text{s.t.}\;L\leq \sum_{j=1}^{n}c_{j}\leq H,\;c_{j}k_{min}\leq x_{j} \leq c_{j}k_{max},\;\; \forall j=1,...,n\). \cite{XIDONAS201760} defined a set of scenarios, indexed by \(t\in T\), that describe the assets' performance, each has an expected return vector \(r_{t}\) and a covariance matrix \(Q_{t}\). They also defined \({v_{t}^{2}}^{*}\) as the minimum variance of a portfolio under scenario \(t\), which is calculated by solving the classical mean-variance PSP for \(Q=Q_{t}\). The final formulation tries to find the optimal solution in the worst-case situation as \(\min_{x\in X,s}\;\text{s.t.}\;x'Q_{t}x\leq (1+s){v_{t}^{2}}^{*},\;\;\forall t\in T,\;L\leq \sum_{j=1}^{n}c_{j}\leq H,\;c_{j}k_{min}\leq x_{j} \leq c_{j}k_{max},\;\; \forall j=1,...,n\), where \(s\) is the relative worst variance aggravation based on the robust solution. 

The risk parity or the equal risk contribution is a new asset allocation strategy in which all of the underlying assets in the portfolio contribute equally to the risk. It has been argued that risk parity results in a superior Sharpe ratio than the mean-variance PSP (see, \eg \cite{demiguel2009optimal}). However, inputs of the risk parity formulations are often subject to uncertainty, which leads to sub-optimal solutions. \cite{demiguel2009optimal} assumed that \(\rho(.)\) is a continuously differentiable convex risk measure, and \(b_{j}\) are the risk budgets assigned by the investor. The risk budgeting problem becomes \(x^{*}=x_{j}\frac{\partial \rho(x)}{\partial x_{j}}=b_{j}\; \forall j,\;x\in X\),
where \(\frac{\partial \rho(x)}{\partial x_{j}}\) is the marginal risk contribution and \(x_{j}\frac{\partial \rho(x)}{\partial x_{j}}\) is the risk contribution of asset \(j\), which has the optimal solution \(x^{*}=\frac{w^{*}}{e'w^{*}}\), where \(w^{*}=\argmin_{w\geq 0}\{\rho(w)-\sum_{j=1}^{n}b_{j}\ln{w_{j}}\}\). \cite{kapsos2018robust} used the variance of the portfolio, which is uncertain and belongs to an uncertainty set \(U_{Q}\), to quantify risk. With that, the robust counterpart of the risk budgeting problem becomes \(\min_{x \in X}\max_{Q\in U_{Q}}\;\; ({x'Qx-\sum_{j=1}^{n}\ln{b_{j}}x_{j}})\),
which is equivalent to \(\min_{w\geq 0}\max_{Q\in U_{Q}}\;\; (w'Qw-\sum_{j=1}^{n}\ln{b_{j}}w_{j})\),
where \(x^{*}=\frac{w^{*}}{e'w^{*}}\) is a normalization of decision variables. \cite{kapsos2018robust} proposed three robust risk budgeting formulations, for which the covariance matrix of assets belongs to; a discrete uncertainty set, a box uncertainty set while the upper bound is a PSD matrix, or a box uncertainty set without restrictions on its bounds. In the last case, the formulation is transformed to a semi-infinite problem that is solvable using an iterative procedure proposed by the authors. 

\subsubsection{Other Extreme Cases of the Mean-Variance}

Worst-case RO is an extreme case, which finds the optimal solution of an optimization problem for the worst possible situation. However, this approach is over-conservative. The goal of reducing the conservatism of RO solutions can be achieved by using other extreme cases than worst-case. \cite{chen2019robust} incorporated set order relations of solutions into a multi-objective mean-variance PSP with an ellipsoidal uncertainty set
to show the relationship between optimization solutions and their efficiency by comparing multiple objective function values. These relations can be interpreted as extreme cases. The first relation, called ``upper set less ordered relation", is the best solution for the worst-case situation, which is equivalent to the robust formulation. The second case is ``lower set less ordered relation" which practically means the best-case solution. Third, ``alternative set less ordered relation", is the intersection of the best-case and the worst-case solutions, \ie \(X_{alternative}=X_{best-case}\cap X_{worst-case}\).   
This study assumed that the distributions of asset returns are normal. \cite{chen2018robust}, however, argued that practical and theoretical evidence shows that the distribution function of asset returns has a fat-tail. Hence, they applied the relation structure of \cite{chen2019robust} and the idea of other extreme cases in PSP without the normality assumption by using the higher moments (skewness and kurtosis) in their formulation. Both \cite{chen2019robust} and \cite{chen2018robust} used a multi-objective particle swarm optimization algorithm to solve other extreme cases of the mean-variance PSP. 

Extreme cases (worst-case, best-case, and  the intersection of the best-case and the worst-case solutions, can be implemented in different market conditions. \cite{bai2019improving} considered different realizations of the uncertain parameters in different market conditions by dividing the market situation into bull market, bear market, and steady market. In the bull market condition, it is assumed that the best-case scenario will happen, hence a best-case formulation (\ie min-min or max-max) is used. Conversely, in the bear market condition, it is assumed that the worst-case scenario will happen, leading to a typical worst-case RO. In the steady market, an alternative scenario, namely the intersection of solutions of the best-case and the worst-case scenarios is assumed to happen. In contrast to \cite{chen2018robust} and \cite{chen2019robust}, \cite{bai2019improving} used a single objective mean-variance PSP.

\subsubsection{Robust Mean-Variance and Regularization}

An important criticism of the classical mean-variance PSP is its weak performance in out-of-sample data due to overfitting. It also has been shown that, for a large number of periods, the classical formulation of the mean-variance PSP amplifies the effects of noise, leading to an unstable and unreliable estimate of decision vectors. To reduce these effects, \cite{dai2019sparse} proposed a \emph{sparse} robust formulation for the mean-variance PSP, which places controls on the asset weights in the portfolio. The process of adding information to solve an ill-posed problem is called \emph{regularization}. \cite{dai2019sparse} defined \(r_{s}=(r_{1s},r_{2s},...,r_{ns})\in \mathbb{R}^{n}\) as a vector of asset returns at time \(s, (s=1,...,S)\). Moreover, \(\mathbb{E}[r_{s}]=\bar{r}\) and \(Q=\mathbb{E}[(r_{s}-\bar{r})(r_{s}-\bar{r})']\) are mean vector and covariance matrix, respectively. The portfolio variance is \(x'Q x=\mathbb{E}[|x'\bar{r}-x'r_{s}|^{2}]=\frac{1}{S}||x'\bar{r} e-Rx||_{2}^{2}\), where \(R\) is a matrix whose \(s^{th}\) row equal to \(r_{s}\). If the expectations are replaced by the sample average, then the model can be expressed as a statistical regression, which takes the form \(\min_{x \in X}\frac{1}{S} ||x'\bar{r} e-Rx||_{2}^{2}\), where \(||.||_{2}\) is the \(l_2\) norm. If the size of \(R\) is large, then it amplifies the effects of noise, leading to an unstable and unreliable estimate of the vector \(x\). To overcome this issue, a regularization is applied in the formulation as \(\min_{x\in X}(\frac{1}{S} ||x'\bar{r} e-Rx||_{2}^{2}+\tau ||x||_{1}^{1})\), where \(\tau\) is the parameter for adjusting the relative importance of the \(l_1\) norm penalty in the objective function. However, this sparse formulation does not consider return as an uncertain parameter.
Consequently, two robust formulations of the mean-variance PSP with box and ellipsoidal uncertainty sets are proposed. The results showed that the sparse robust mean-variance PSP has better out-of-sample performance than other mean-variance formulations.
\cite{lee2020sparse} extended the same concept to a robust sparse cardinality-constrained mean-variance PSP with ellipsoidal uncertainty set and \(l_{2}\) norm regularization to achieve a better control over decision variables. This formulation results in a non-convex NP-hard problem. Hence, a relaxation to a SDP problem is proposed to make it more tractable.

Alternatively, it is possible to prevent the negative impact of noisy inputs by adding restrictions on the estimated parameters instead of restricting the decision variables. \cite{plachel2019unified} used the restricted estimators method with a box uncertainty set to derive a robust regularized minimum variance PSP based on the decomposition of covariance matrix proposed by \cite{laloux1999noise}. The proposed formulation was tested with the three major turmoils of the financial market (Black Monday, the Dotcom Bubble, and the Financial Crisis) and the results showed that the joint problem regularization and robustification outperforms the classical non-robust minimum variance and the non-regularized minimum variance PSP.

\subsubsection{Robust Estimators for the Mean-Variance PSP}

The classical mean-variance PSP is based on the Gaussian distribution assumption of asset returns. Based on historical evidence, \cite{lauprete2003robust} showed that the returns of assets follow a heavy-tail distribution. Given that the uncertainty associated with the deviation of actual distribution functions from theoretical distribution functions might lead to sub-optimal solutions, they proposed a robust estimation that immunizes the estimators against uncertainty. 
\cite{demiguel2009portfolio} used two types of robust estimators (M-estimator and S-estimator) in the mean-variance PSP. M-estimator and S-estimator are based on convex symmetric and Tukey’s bi-weight loss functions, respectively. The S-estimator has the advantage of not being sensitive to data scaling.
\cite{demiguel2009portfolio} analyzed the sensitivity of M-portfolios and S-portfolios' (corresponding to M-estimator and S-estimator, respectively) weights with respect to the changes in the distribution of the asset returns. Results showed that these formulations are more robust than the traditional mean-variance PSP.

\subsubsection{Experimental Analysis of the Robust Mean-Variance}

\cite{kim2013robust} identified a gap in the literature about the experimental evidence of the robust PSP. They analyzed the robust mean-variance PSP with box and ellipsoidal uncertainty sets. Results showed that weights in the robust mean-variance PSP align with assets having a higher correlation with the Fama-French three factors model which bets on fundamental factors of assets. Interested readers are refereed to \cite{fama2021common} for more detail about Fama-French three factors model. \cite{kim2014deciphering} also concluded that robust solutions of the mean-variance PSP depend on fundamental factors movements. In another analysis, \cite{kim2013composition} showed that robust equity mean-variance portfolios have four advantageous characteristics compared to non-robust mean-variance PSPs: (1) fewer stocks, (2) less exposure to each stock (the amount of money that the investor could lose on an investment), (3) higher portfolio beta, and (4) large negative correlation between weight and stock beta. \cite{kim2018robust} concluded that the robust mean-variance PSP leads to the most efficient investment strategies that allocate risk efficiently. \cite{kim2015focusing} also illustrated that the robust approach is the best method for formulating the mean-variance PSP while the market switches between multiple states.

Similarly, \cite{schottle2009robustness} analyzed the Markowitz efficient frontier of robust mean-variance PSP with ellipsoidal and joint ellipsoidal uncertainty sets. They showed that the efficient frontiers of both robust formulations are exactly matched with the efficient frontier of the classical mean-variance PSP up to some level of risk. This means that the classical mean-variance PSP is already robust without applying RO methods. However, the robust mean-variance formulation identifies the unreliable upper part of the efficient frontier. 

Recently, \cite{yin2021practical} proposed a practical guide to robust portfolio optimization based on mean-variance formulations. They assumed that asset returns are uncertain and belong to either box or an ellipsoidal uncertainty sets. By using practical examples, they showed that the robust mean-variance PSP with an ellipsoidal uncertainty set provides a more robust formulation than its corresponding problem with a box uncertainty set.

\subsection{Robust Mean Absolute Deviation}

\cite{konno1991mean} argued that calculating the covariance matrix in large mean-variance PSPs is a challenging task. Hence, they proposed the mean absolute deviation as an alternative volatility-based risk measure that reduces the computational complexity of the covariance matrix. The MAD PSP is formulated as \(\min_{x\in X}\frac{1}{S}\sum_{s=1}^{S}|\sum_{j=1}^{n}(r_{js}-r_{j})x_{j}|\;\text{s.t.}\;\sum_{j=1}^{n}r_{j}x_{j}\geq E W_{0}\). This formulation can be transformed to an LP as \(\min_{x \in X,y}\;\; \frac{1}{S}\sum_{s=1}^{S}y_{s}\;\text{s.t.}\; y_{s}+\sum_{j=1}^{n}(r_{js}-r_{j})x_{j}\geq 0,\;\forall s,\;y_{s}-\sum_{j=1}^{n}(r_{js}-r_{j})x_{j}\geq 0,\;\forall s,\;\sum_{j=1}^{n}r_{j}x_{j}\geq E W_{0}\). Besides being reformable as an LP, the MAD PSP has another advantage over the mean-variance PSP of not requiring the normality assumption for asset returns. However, MAD penalizes both positive and negative deviations equally, though positive deviations are desirable by investors. Moreover, in the classical MAD PSP, future asset returns are assumed to be known with certainty.  

To handle the uncertainty of asset returns in the MAD PSP, \cite{moon2011robust} proposed a robust MAD PSP with a budget uncertainty set. However, \cite{li2016portfolio} suggested that classical uncertainty sets do not capture the asymmetry in asset returns and, instead, proposed a robust MAD PSP with the asymmetric uncertainty set first introduced by \cite{chen2007robust}.  \cite{ghahtarani2018robust} developed a robust PSP based on m-MAD, a downside risk measure proposed by \cite{michalowski2001extending} that penalizes only negative deviations. \cite{chen2011tight} proposed an alternative robust downside risk measure, referred to as \emph{lower partial moment} (LPM), and used it, along with a moment-based ambiguity set, for single and multi-stage robust PSP that uses an $S$-shape value function. An important advantage of robust LPM over robust m-MAD is that the former can be used also to develop robust VaR/CVaR formulations with moment-based ambiguity sets. To avoid the over-conservatism of worst-case approaches, \cite{xidonas2017robust} employed a robust min-max regret approach in a multi-objective PSP. The objectives to be optimized are expected asset returns and MAD. The proposed approach results in solutions that do not have to be safe according to the worst realization of the parameters, but to the relevant optimum of each scenario.

\subsection{Factor-Based Portfolio Models}

Factor-based models are financial models that incorporate factors (macroeconomic, fundamental, and statistical) to determine the market equilibrium and calculate the required rate of return. \cite{goldfarb2003robust} developed a robust factor-based model for a PSP where uncertainty is considered by its sources, namely fundamental factors. The basic formulation of a factor-based model is \(r=\mu+V^{'}f+\epsilon\), where \(\mu \in \mathbb{R}^{n}\) is the mean returns vector, \(f\sim N(0,F)\) is the vector of the factors that drive the market, \(V \in R^{m\times n}\) is the matrix of the factor loading of \(n\) assets, and \(\epsilon \sim N(0,D)\) is the vector of the residual returns. Uncertain parameters are the mean return, the factor loading, and the covariance of residuals that belong to uncertainty sets with upper and lower bounds. \cite{goldfarb2003robust} defined uncertainty sets for these parameters as \(U_{d}=\{D: D=diag(d),\;\; d_{j}\in [ \,\underline{d}_{j},\bar{d}_{j}],\; \forall j\)\}, \(U_{\nu}=\{V: V=V_{0}+W,\; ||W_{j}||_{g}\leq\rho_{j},\;\forall j\}\), and \(U_{m}=\{\mu: \mu=\mu_{0}+\xi,\;\;\;|\xi_{j}|\leq\gamma_{j},\;\forall j\}\), where \(W_{j}\) is the \(jth\) column of \(W\) and \(||w||_{g}=\sqrt{w'Gw}\) is an elliptic norm of \(w\) with respect to \(G\). The return on a portfolio \(x\) is given by \(r_{x}=r'x=\mu'x+f'Vx+\epsilon'x\sim N(x'\mu,x'(V'FV+D)x)\). Both \(f\) and \(\epsilon\) are assumed to follow normal distributions, thus \(r_{x}\) also follows a normal distribution. The robust factor-based model is developed based on two alternative assumptions. First, uncertainty in the mean is independent of the uncertainty in the covariance matrix of returns, which leads to a SOCP. Second, uncertainty in the mean depends on the uncertainty in the covariance matrix of the returns, which results in a SDP formulation for the worst-case VaR. It should be noted that the uncertainty sets in the robust factor-based models of \cite{goldfarb2003robust} are separable, leading to two important drawbacks: the results are conservative, and the robust portfolio constructed is not well diversified. Alternatively, \cite{lu2006new} and \cite{lu2011robust} proposed robust factor-based models with a joint ellipsoidal uncertainty set that can be reformulated as a tractable cone programming problem. Additionally, 
\cite{ling2012robust} developed a robust factor-based model with joint marginal ellipsoidal uncertainty sets and options to hedge risks that generates robust portfolios with good wealth growth rates even if an extreme event occurs.

An important input to factor-based models is the ``factor exposure", which measures the reaction of factor-based models to risk factors. \cite{kim2014robust} argued that factor-based models are not robust against the uncertainty of risk factors such as macroeconomic factors. They proposed a robust factor-based model with an ellipsoidal uncertainty set that is robust against uncertainty and has the desired level of dependency on factor movements. This model manages the total portfolio risk by defining a robustness measure and a constraint that restricts the factor exposure of robust portfolios.
Another evidence to support the use of robust factor-based models comes from \cite{lutgens2010robust}. They compared the Capital Asset Pricing Model (CAPM), the international CAPM, the international Fama, and the French factor-based models and showed that robust portfolios of factor models lead to better diversified portfolios.

\subsection{Robust Utility Function PSP}

Most PSPs are based on the return-risk trade-off concept. However, financial decisions might be made based on a utility function. \cite{popescu2007robust} developed a robust PSP for the expected utility function (the utility that an entity or aggregate economy is expected to reach under any number of circumstances) where the distribution function of asset returns is partially known and belongs to an ambiguity set with predetermined mean vector and covariance matrix.
\cite{natarajan2010tractable} also proposed a less-complex robust formulation of the expected utility function PSP that uses a piecewise-linear concave function to model the investor's utility. Besides the ambiguity set of \cite{popescu2007robust}, \cite{natarajan2010tractable} considered the case in which the mean vector and covariance matrix of uncertain parameters belong to box uncertainty sets. \cite{ma2008robust} incorporated a robust factor model with a concave-convex utility function to seize the advantages of both approaches. They assumed that the mean returns vector, the factor loading covariance, and the residual covariance matrix are uncertain and belong to uncertainty intervals. The robust counterpart turned out to be a parametric quadratic programming problem that can be solved explicitly. \cite{biagini2017robust} proposed a min-max robust utility function for Merton problem. Merton's portfolio problem is a well-known PSP problem where the investor must choose how much to consume and how allocate the remaining wealth between risky assets and a risk-free asset to maximize expected utility. An ellipsoidal uncertainty set is assumed to contain the drift from a compact values volatility realization.

\section{Robust PSPs with Quantile-based Risk Measures}\label{quantile}

This section reviews robust quantile-based PSPs, which include PSPs based on the Value at Risk (VaR), Conditional Value at Risk (CVaR), and their extensions with worst-case RO, relative RO and distributionally robust optimization (DRO) methods. Furthermore, the relationship between uncertainty sets and risk measures, application of soft robust formulation with risk measures, worst-case CVaR and its relationship with the uniform investment strategy, and robust arbitrage pricing theory with worst-case CVaR are also discussed.

VaR is the maximum loss at a specific confidence level. In other words, VaR is the quantile of a loss distribution function, which is neither a convex nor coherent risk measure. A coherent risk measure is a function that satisfies the properties of monotonicity, sub-additivity, homogeneity, and translational invariance which provide computational advantages for a risk measure (see \cite{artzner1999coherent} for details about coherent risk measures). CVaR is a coherent risk measure that denotes expected loss greater than VaR for a specific confidence level. 
Let \(f(x,r)\) be a loss function. For a given confidence level \(\beta\), the Value at Risk is defined as \(VaR_{\beta}(x)=\min\{\alpha \in \mathbb{R}: \Psi(x,\alpha)\geq \beta\}\),
where \(\Psi(x,\alpha)=\int_{f(x,r)\leq\alpha}p(r)dr\).
Conditional Value at Risk is the expected loss that exceeds \(VaR_{\beta}(x)\), mathematically defined as \(CVaR_{\beta}(x)=\frac{1}{1-\beta}\int_{f(x,r)\geq VaR_{\beta}(x)}f(x,r)p(r)dr\).
\cite{rockafellar2000optimization} proved that CVaR can be formulated as an optimization problem by defining an auxiliary function \(F_{\beta}(x,\alpha)=\alpha+\frac{1}{1-\beta}\int_{y\in R^{m}}[f(x,r)-\alpha]^{+}p(r)dr\),
where \([.]^{+}=\max\{.,0\}\) and \(CVaR_{\beta}(x)=\min_{\alpha \in R}F_{\beta}(x,\alpha)\). They also proved that CVaR can be reformulated as an LP
when using discrete scenarios for asset returns. 
An important input to these formulations is \(p(.)\), which is often not known or only partially known. Considering the ambiguity of \(p(.)\) leads to worst-case VaR and CVaR formulations.

\subsection{Worst-Case VaR and CVaR}

\cite{ghaoui2003worst} were the first to propose a tractable reformulation for the worst-case VaR, defined as \(VaR_{p}(x)=\min{\alpha}\;\text{s.t.}\; \sup\{Prob{\Psi(x,\alpha)\geq \beta\}}\leq \epsilon\), where \(VaR_{p}^{optimum}=\min{VaR_{p}(x)}\;s.t.\;x\in X\).
The distribution function of asset returns is assumed to be partially known and belongs to one of the four moment-based ambiguity sets: 1) the first two moments (mean vector \((\hat{r})\) and covariance matrix \((\Sigma)\)) of the loss distribution function are known and fixed. 2) the moments \((\Sigma, \hat{r})\) of the loss distribution function are known to belong to the convex set, , assuming that there is a point in \(U\) such that \(\Sigma\succ0\). By introducing \(U_{+}:=\{(\Sigma,\hat{r})\in U| \Gamma\succ 0\}\), the worst-case VaR for this case is formulated as \(VaR_{p}(x)=\sup{-r'x}\;\text{s.t.}\;(\Sigma,\hat{r})\in U_{+}\). 3) polytopic uncertainty set defined as the convex hull of the vertices \((\hat{r_{1}},\Sigma_{1}),...,(\hat{r_{l}},\Sigma_{l})\).
The polytope uncertainty set \(U\) is then constructed as \(U=U_{r}\times U_{\Sigma}\), where \(U_{r}=Co\{\hat{r_{1}},...\hat{r_{l}}\}\) and \(U_{\Sigma}=Co\{\Sigma_{1},...,\Sigma_{l}\}\). By assuming that \(\Sigma_{i}\succ 0,\;\;i=1,...,l\), the worst-case VaR is formulated as \(VaR_{p}(x)=k(\varepsilon)\sqrt{\max_{\Sigma\in U_{\Sigma}}x' \Sigma x}-\min_{\hat{r}\in U_{r}}{\hat{r}'x}=\max_{1\leq i\leq l}{k(\varepsilon)||\Sigma_{i}^{1/2}||_{2}-\min_{1\leq i \leq l}\hat{r'}_{i}x}\),
where \(k(\varepsilon)=\sqrt{\frac{1-\varepsilon}{\varepsilon}}\). \cite{ghaoui2003worst} showed that this formulation can be transformed to a SOCP model. 4) componentwise bounds for moments. \cite{ghaoui2003worst} also considered the worst-case VaR when the return of assets in the loss function is based on the factor model \(r=Vf+\epsilon\), where \(f\) is an $m$-vector of random factors, \(\epsilon\) is the residual (unexplained) return, and \(V\) is an \(n\times m\) matrix of sensitivities of the returns. The covariance matrix of returns is stated as \(\Sigma=D+VSV'\), where $D$ is the diagonal covariance matrix of residuals and $S$ is the covariance matrix of factors. Two cases of parameter certainty are considered: uncertainty in the factor's mean and covariance matrix, and uncertainty in the sensitivity matirx. 
In contrast to \cite{ghaoui2003worst}, the factor model of \cite{goldfarb2003robust} assumed that uncertainty in the mean is independent from that of the covariance matrix, leading the expected value of error term of the factor model to be equal to zero. This uncertainty structure leads to a SOCP reformulation, compared to the SDP reformulation of \cite{ghaoui2003worst}.

It is argued that the worst-case VaR is unrealistic and conservative. Therefore, a way to enforce the worst-case probability distribution to some level of smoothness was proposed by adding a relative entropy constraint (\ie KL divergence) with respect to a given ``reference" probability distribution. Whereas \cite{ghaoui2003worst} assumed that the return of assets follows a Gaussian distribution, \cite{belhajjam2017robust} argued that the distribution function of return is asymmetric. Hence, extreme returns occur more frequently than would be under the normal distribution. Hence, they proposed a multivariate extreme Value at Risk (MEVaR) formula based on a multivariate minimum return that 
considers extremums of returns, \ie the lowest and highest daily returns. 
Since there is no guarantee that uncertain parameters belong to a symmetric uncertainty set, \cite{natarajan2008incorporating} applied the asymmetric uncertainty set introduced by \cite{chen2007robust} to develop a worst-case VaR measure. Results show that Asymmetry-Robust VaR (ARVaR) is an approximation of CVaR. Similar to \cite{ghaoui2003worst}, \cite{natarajan2008incorporating} assumed that asset returns follow a factor model. Moreover, an asymmetric uncertainty set for the worst-case VaR leads to a tractable second-order cone program. Another less complex method to consider asymmetric uncertainty is to use interval random uncertainty sets. \cite{chen2011worst} developed a worst-case VaR assuming that the expected vector and covariance matrix of the returns are uncertain and belong to interval random uncertainty sets. 

\cite {huang2007robust} demonstrated that the exit time of investment (or the investment horizon) which is traditionally assumed to be deterministic, can, in reality, depend on market conditions. Consequently, they considered a conditional distribution function of the rate of return based on different exit times instead of the unconditional distribution function previously used in \cite{ghaoui2003worst}. Three robust portfolio formulations were proposed: 1) a portfolio formulation with componentwise uncertainty on moments of the conditional distribution function of exit time. 2) a portfolio formulation with semi-ellipsoidal uncertainty set on exit time. 3) moments of the conditional distribution function of exit time belonging to a polytope uncertainty set for each exit time. \cite{huang2008portfolio} also assumed that the density function of exit time is only known to belong to an ambiguity set that covers all possible exit scenarios. They developed two formulations: a worst-case VaR with no information about exit time, and a formulation with partial information about exit time.

\cite{kelly1956new} proposed an investment strategy in the financial market (known as \emph{Kelly Strategy}), which maximizes an expected portfolio growth rate. From a mathematical perspective, implementing the Kelly strategy is synonymous with solving a multi-period investment strategy, making it amenable to robust approaches for handling uncertainty. \cite{rujeerapaiboon2016robust} considered Kelly's strategy under return uncertainty and proposed a formulation that includes the constraint \(\mathbb{P}(total\;portfolio\;return\; \geq \;\gamma)\geq 1-\epsilon\), where \(\gamma\) is an expected total portfolio return, and \(1-\epsilon\) is the confidence level. This chance constraint is, simply, the definition of VaR. In this formulation, the distribution function of asset returns is assumed to be uncertain and belongs to the class of moment-based ambiguity set introduced in \cite{delage2010distributionally}. This ambiguity set leads to a SDP formulation for the worst-case VaR. 

As mentioned earlier, VaR has a high computational complexity since it is not convex. \cite{zhu2009worst} proposed a PSP that maximizes the worst-case CVaR, defined as \(\sup_{\pi \in P}CVaR_{\beta}(x)\), with three cases of uncertainty set for the probabilities of discrete return scenarios: a mixture distribution, a box uncertainty set, and an ellipsoidal uncertainty set. The last case led to a SOCP, whereas the first two cases resulted in LPs. A mixture distribution \((P_{M})\) is defined as \(\pi\in P_{M}=\{\sum_{i=1}^{I}\eta_{i}p^{i}(.): \sum_{i=1}^{I}\eta_{i}=1,\;\; \eta_{i}\geq 0,\;\;i=1,...,I\}\), which leads to: \(WCVaR_{\beta}(x)=\min_{\alpha\in \mathbb{R}}\max_{i\in l}F_{\beta}^{i}(x,\alpha)\), where \(l=[1,...,I]\). The box uncertainty set for probability distribution is defined as \(\pi\in P_{\pi}^{\beta}=\{\pi:\; \pi=\pi^{0}+\zeta,\; e^{'}\zeta=0,\; \underline{\zeta}\leq\zeta\leq\bar{\zeta}\}\), whereas the ellipsoidal uncertainty set for probability distribution function is defiend as \(\pi\in P_{\pi}=\{\pi: \pi=\pi^{0}+A\zeta,\; e'A\zeta=0,\; \pi^{0}+A\zeta\geq0\; ||\zeta||\leq 1\}\), where \(||\zeta||=\sqrt{\zeta' \zeta}\) and \(\pi^{0}\) is the nominal distribution. \cite{doan2015robustness} extended the worst-case CVaR formulation of \cite{zhu2009worst} by proposing a data-driven approach to construct a class of distributions for asset returns, known as Fréchet distributions, that leads to less conservative solutions than the worst-case CVaR. Moreover, \cite{hasuike2018investor} incorporated the arbitrage pricing theory (APT) model, which is a multi-factor model, in a bi-objective PSP that aims at maximizing the expected return and minimizing the worst-case CVaR of a portfolio. \cite{ghahtarani2018development} proposed a robust CVaR formulation by considering the uncertainty of the return distribution's parameters. 
They proposed a robust mean-CVaR PSP with a chance constraint when asset returns follow a Gaussian distribution with uncertain moments. \cite{hellmich2011efficient}, in contrast, developed a worst-case CVaR model with asset returns that follows a heavy-tail multivariate generalized hyperbolic distribution. Their formulation can also capture the asymmetrical nature of asset returns. 

One way to alleviate the over-conservatism of the worst-case VaR/CVaR solutions is to use a data-driven joint ellipsoidal uncertainty set in which the first two moments of the distribution function of asset returns are in an ellipsoid norm. \cite{lotfi2018robust} proposed an algorithm for constructing data-driven ambiguity sets based on an optimization model to find the centers of joint ellipsoidal uncertainty sets. In another attempt, \cite{liu2019closed} used the data-driven moment-based ambiguity set introduced in \cite{delage2010distributionally} to propose a worst-case CVaR in both single and multi-period PSPs. In this formulation, for each period there is a separate ambiguity set. They demonstrated that a robust counterpart of the multi-period mean-CVaR PSP can be solved as a sequence of optimization problems based on an adaptive robust formulation. \cite{kang2019data} argued that the ambiguity set of \cite{delage2010distributionally} leads to solutions that are too conservative. Therefore, they altered it by adding a zero-net adjustment constraint. \cite{huang2021sparse} proposed a distributionally robust mean-CVaR PSP with a moment-based ambiguity set. Besides DRO, they used an \(l_{1}\) norm to limit the weights (decision variables) of the, so called, sparse PSP to limit the impact of noisy data. Results provide evidence that a sparce mean-CVaR PSP has better performance than a non-sparce formulation with respect to net portfolio return, Sharpe ratio, and cumulative return. Moreover, \cite{zhao2021robust} formulated a cardinality-constrained rebalancing worst-case CVaR with a moment-based ambiguity set. The proposed formulation enhances the portfolio diversification.

\cite{huang2010portfolio} claimed that investors usually do not want to pay the price of full robustness to protect their portfolios against the worst possible scenario. In an uncertain environment, investors may rather choose a strategy that avoids falling behind their competitors. According to this point of view, for each choice of decision variables and each scenario, the decision-maker compares the resulted objective value to the optimal value obtained under model uncertainty described by the scenario. The difference or the ratio of these two values is a regret measure. To minimize these regrets measures, \cite{huang2010portfolio} developed a relative CVaR formulation, mathematically described as \(RCVaR_{\alpha}(x)=\sup_{\pi\in P}\{CVaR_{\alpha}(x,\pi)-CVaR_{\alpha}(z^{*}(\pi),\pi)\}\),
where \(z^{*}(\pi)=\argmin_{z\in X}CVaR_{\alpha}(z,\pi)\). However, since the true distribution (\(\pi\)) is not known, decision-makers try to make the relative CVaR as small as possible by considering all possible \(\pi\) values. Consequently, a finite number of forecasts for the distribution function of asset returns is considered. Results showed that the relative CVaR is less conservative than the worst-case CVaR for optimal portfolio return. 
Alternatively, \cite{yu2017incorporating} proposed a relative CVaR and a worst-case CVaR by adjusting the required return from a fixed rate to a floating rate that changes according to market dynamics. Moreover, the formulation was extended by allowing short sale and adding a transaction cost constraint. Results showed that a relative CVaR yields slightly higher realized returns, lower trading costs, and better portfolio diversification than its corresponding worst-case CVaR model when the required return is fixed. Additionally, the out-of-sample performance of floating-return models compared to fixed-rate models is significantly better during periods when a market recovers from a financial crisis. Finally, robust floating-return models have a better asset allocation, save transaction costs, and attribute to superior profitability. \cite{benati2021relative} proposed a model that minimizes the maximum regret on the expected returns while the conditional value-at-risk is bounded under different scenario settings. To solve this problem, a cutting plane approach was proposed.

An investment strategy that is widely used in financial markets is the \emph{uniform investment strategy} or \emph{\(1/N\) rule}, which divides the budget among assets equally. \cite{pflug20121} demonstrated that the uniform investment strategy is the best strategy for investment under uncertainty. They proposed robust mean-CVaR and mean-variance PSPs where the distribution function of asset returns is uncertain and belongs to a \emph{Kantorovich} or \emph{Wasserstein} metric-based ambiguity set. Results showed that when the size of the Wasserstein ambiguity set is infinity, solutions of the robust PSPs are equal to the uniform investment strategy. Hence, the optimal investment strategy in a high ambiguity situation is the uniform investment or \(1/N\) rule. However, \cite{pflug20121} assumed that all assets are subject to uncertainty though it is possible to use fixed-income assets with no ambiguity or uncertainty in the portfolio. Therefore, \cite{pacc2018robust} extended the robust uniform strategy of \cite{pflug20121} by considering both ambiguous and unambiguous assets. They showed that by increasing the ambiguity level, measured by the radius of the ambiguity set, the optimal portfolio tends to use equal weights for all assets. Also, high levels of ambiguity result in portfolios that avoid ambiguous assets and favor unambiguous assets. 

Finally, \cite{natarajan2009constructing} established the relationship between risk measures and uncertainty sets. They showed that using an ellipsoidal uncertainty set for asset returns corresponds to the classical mean-variance PSP, whereas the CVaR formulation results from using a special polyhedral uncertainty set. As discussed by \cite{ben2010soft}, in soft robust formulations, a penalty function is introduced such that if uncertain parameters fluctuate in the uncertainty set, the penalty function equals zero. Otherwise, the penalty function takes a positive value. \cite{recchia2014robust} proved that the definition of a convex risk measure is also based on a penalty function that is called norm-portfolio models, where using \(l_{\infty}\), \(l_{1}\), and $D$-norm result in an LP for a norm-portfolio model. On the other hand, using a euclidean norm results in a SOCP, whereas applying a $D$-norm, proposed by \cite{bertsimas2004robust}, for a penalty function with specific parameters leads to the CVaR formulation.

\subsubsection{Worst-Case CVaR with Copula}

Classical multivariate distribution functions make the worst-case CVaR computationally complex. One way to address this issue is to use copulas instead of multivariate distribution functions for asset returns. Copulas are multivariate distribution functions whose one-dimensional margins are uniformly distributed on a closed interval \([0,1]\). One-dimensional margins of copulas can be replaced by univariate cumulative distributions of random variables. Hence, copulas consider the dependency between marginal distributions of random variables instead of focusing directly on dependency between random variables themselves. This characteristic makes them more flexible than standard distributions, and also an interesting candidate for the distribution function of the rate of return in the worst-case CVaR. 

\cite{kakouris2014robust} used Archimedean copulas to propose worst-case CVaR PSPs that avoids the shortcomings of worst-case CVaR PSPs based on a Gaussian distribution, which is a symmetric distribution for asset returns. There are three Archimedean copulas: the Clayton copula, the Gumbel copula, and the Frank copula. \cite{kakouris2014robust} used a heuristic method to estimate copulas' parameters in the context of a multi-asset PSP. However, simulating data from three Archimedian copulas has computational challenges.  
On the other hand, \cite{han2017dynamic} claimed that the formulation of \cite{kakouris2014robust} is static, making it unable to deal with the dynamic nature of the financial market. They, instead, proposed a dynamic robust PSP with Archimedean copulas by using dynamic conditional correlation (DCC) copulas and copula-GARCH model to forecast the worst-case CVaR of bi-variate portfolios. Results show that dynamic worst-case CVaR models can put more weight on assets with lower volatility, which leads to a less aggressive trading strategy.

CVaR calculates the expected loss based on just one confidence level. However, decision-makers might prefer different confidence levels based on their risk attitude. One way to increase the flexibility of CVaR related to decision-makers' risk attitude is to use Mixed-CVaR and Mixed Deviation-CVaR. These mixed risk measures combine CVaRs with different confidence levels. \cite{goel2019robust} proposed robust Mixed-CVaR and Mixed Deviation-CVaR Stable Tail-Adjusted Return Ratio (STARR), which is the portfolio return minus the risk-free rate of return divided by the expected tail loss (at a specific confidence level). Finally, a mixture copula set was used to consider distribution ambiguity, which resulted in an LP.

\subsection{Robust Mean-CVaR/Shortfall PSP}

Besides the worst-case VaR and CVaR, some researchers developed robust mean-CVaR PSPs where the distribution function of the loss function is assumed to be deterministic while returns of assets or weights of the mixture distribution function of the rate of return are uncertain. Thus, classical uncertainty sets are used to develop robust mean-CVaR PSPs. 
\cite{quaranta2008robust} proposed a robust mean-CVaR with a box uncertainty set that leads to an LP. \cite{kara2019stability} also proposed a robust mean-CVaR PSP with a parallelepiped uncertainty set, developed by \cite{ozmen2011rcmars}. The parallelepiped uncertainty set is practically a box uncertainty set while its elements are a convex hull of canonical vertices of an uncertain matrix. Elements of this uncertainty set are founded by the Cartesian product of uncertain intervals. An advantage of a parallelepiped uncertainty set over a box uncertainty set is that the lengths of intervals may vary among each other. Moreover, instead of a single price, it is possible to consider multiple and flexible varying prices of assets and also take into account likewise flexible returns.
To reduce the conservatism of solutions of a robust mean-CVaR with a box uncertainty set, \cite{guastaroba2011investigating} developed a robust mean-CVaR with ellipsoidal and budget uncertainty sets, which lead to a SOCP and an LP, respectively. Besides the uncertainty of parameters, a mean-CVaR PSP has a multi-objective characteristic as it maximizes the expected return while minimizing the risk (CVaR). Then, a multi-objective formulation can capture the multiple-criteria nature of this problem. \cite{rezaie2015ideal} developed a robust bi-objective mean-CVaR PSP with a budget uncertainty set. An ideal and anti-ideal compromise programming approach was used to solve the proposed problem. This method seeks an answer as close as possible to the ideal value and as far as possible from the anti-ideal value of each objective. Ideal and anti-ideal values reflect investors' perspectives of the real world.

Another development of a robust mean-CVaR is based on mixture distribution functions. There are three reasons for using a mixture distribution function for asset returns. First, it is a combination of multiple distribution functions, thus enabling different market conditions with different distribution functions to be considered. Moreover, it replaces the estimation of the distribution function by a calculation of the distribution weights in a mixture distribution function. Finally, since any distribution function can be simulated by using a mixture of Gaussian distribution functions, a mixture distribution function has high flexibility. \cite{zhu2014portfolio} used a mixture distribution function for asset returns to propose a robust mean-CVaR PSP. The uncertainty in their formulation is about the weights of distribution functions. For considering the uncertainty, ellipsoidal and box uncertainty sets were used. The former leads to a SOCP and the latter results in an LP. 

Shortfall is also a quantile risk measure from the family of VaR and CVaR, introduced by \cite{bertsimas2004shortfall}. Shortfall measures how great an expected loss will be if a portfolio return drops below the \(\alpha\)-quantile of its distribution. Mathematically, it is defined as \(S_\alpha=\mathbb{E}[r'x]-\mathbb{E}[r'x\;|\;r'x \leq q_{\alpha}(r'x)],\;\alpha \in (0,1)\), where \(q_{\alpha}\) is the $\alpha$-quantile of the distribution of random portfolio return. Like CVaR, shortfall can be reformulated as an LP while asset returns are subject to uncertainty. Later, \cite{pachamanova2006handling} developed a robust shortfall with an ellipsoidal uncertainty set, which can be reformulated as a SOCP. Their results showed that a robust shortfall PSP outperforms its nominal problem in the presence of uncertainty in terms of both return and risk.
Another quantile-based measure is the conditional expectation type reward–risk performance measure developed by \cite{ortobelli2019use}. This performance measure captures the portfolio’s distributional behaviour on the tails. \cite{kouaissah2021robust} proposed a robust conditional expectation formulation where the asset returns are uncertain and belong to an ellipsoidal uncertainty set. Results of this robust formulation demonstrated better out-of-sample performance than its nominal counterpart.

\section{Multi-Period PSP} \label{multi-period}

Active strategies which involve ongoing buying and selling of assets are preferred by many investors. With an active strategy, investors continuously re-balance their portfolios by solving multi-period PSPs. In this section, we review applications of RO in this class of problems.

\subsection{Robust Multi-Period PSP}

\cite{dantzig1993multi} proposed one of the most popular multi-period PSPs. Three types of decision variables are used in their formulation: \(x_{j}^{s}, y_{j}^{s}\) and \(z_{j}^{s}\), denoting, respectively, the amounts of asset \(j\) at period \(s\) the investors hold, buy and sell. There are \(n\) risky assets and one risk-free asset. The problem is formulated as \(\max \sum_{j=1}^{n+1} r_{j}^{S} x_{j}^{S}\;\text{s.t.}\;x_{j}^{s}=r_{j}^{s-1}x_{j}^{s-1}-y_{j}^{s}+z_{j}^{s},\;\forall j,s,\;x_{n+1}^{s}=r_{n+1}^{s-1}x_{n+1}^{s-1}+\sum_{j=1}^{n}(1-\Delta_{j}^{s})y_{j}^{s}-\sum_{j=1}^{n}(1+\nu_{j}^{s})z_{j}^{s},\;y_{j}^{s}\geqslant0,\;z_{j}^{s}\geqslant0,\;\forall j,s,\;x_{j}^{s}\geqslant0\), where the objective function maximizes the total wealth at the final period. The first constraint is for risky assets balancing, ensuring that the amount of risky assets held at period \(s\) equals the amount of assets carried forward from the previous period in addition to the net effect of transactions in the current period. The second constraint is for risk-free asset balancing, where \((1-\Delta_{j}^{s})y_{j}^{s}\) is the amount of cash investors receive from selling asset \(j\) at the beginning of the period \(s\), whereas \((1-\nu_{j}^{s})z_{j}^{s}\) is the cash investors use to buy asset \(j\) at the beginning of period \(s\). The uncertain parameters in this formulation are asset returns at each period. \cite{ben2000robust} reformulated this multi-period PSP by defining cumulative asset returns \(R_{j}^{s}=r_{j}^{0}r_{j}^{s}...r_{j}^{s-1}\), which become the new uncertain parameters. By considering these cumulative returns, \cite{ben2000robust} defined new variables for their formulation as \(\xi_{j}^{s}=\frac{x_{j}^{s}}{R_{j}^{s}}\), \(\eta_{j}^{s}=\frac{y_{j}^{s}}{R_{j}^{s}}\), and \(\zeta_{j}^{s}=\frac{z_{j}^{s}}{R_{j}^{s}}\). The final formulation becomes \(\max \sum_{j=1}^{n+1} R_{j}^{S+1} \xi_{j}^{S}\;\text{s.t.}\;\xi_{j}^{s}=\xi_{j}^{s-1}-\eta_{j}^{s}+\zeta_{j}^{s},\; \forall j,s,\;\xi_{n+1}^{s}=\xi_{n+1}^{s-1}+\sum_{j=1}^{n}A_{j}^{s}\eta_{j}^{s}-\sum_{j=1}^{n}B_{j}^{s}\zeta_{j}^{s},\;\forall s,\;\eta_{j}^{s}\geqslant0,\;\;\;\;\zeta_{j}^{s}\geqslant0,\;\forall j,s,\;\xi_{j}^{s}\geqslant0,\;\forall j,s\), where \(A_{j}^{s}=(1-\Delta_{j}^{s})\frac{R_{j}^{s}}{R_{n+1}^{s}}\), and \(B_{j}^{s}=(1-\nu_{j}^{s})\frac{R_{j}^{s}}{R_{n+1}^{s}}\). Both SP and RO were applied with the last nominal formulation. Interestingly, the RO problem was shown to be less complex than the SP one. An ellipsoidal uncertainty set was used in the robust problem, leading to a SOCP. Alternatively, \cite{bertsimas2008robust} used a \(D\)-norm to define the uncertainty set for cumulative asset returns, thus leading to a tractable LP reformulation. To make the problem more appealing for practitioners and to increase its robustness against market volatility,
\cite{marzban2015developing} included American options in the robust
formulation of \cite{bertsimas2008robust}. However, this change led to more
conservative solutions in comparison to those of \cite{ben2000robust} and \cite{bertsimas2008robust}. 

\cite{fernandes2016adaptive} added a loss function with a predetermined threshold as a constraint to the formulation of \cite{dantzig1993multi}, leading to a problem with a terminal wealth objective that requires a one-step-ahead asset return forecast as an input. A linear combination of chosen predictors is employed as a mixed-signals model that uses the last specific number of trading periods to forecast one-step ahead returns. A polyhedral set, constructed as the convex hull of the observed returns, is used as a data-driven uncertainty set. The proposed loss constraints adaptively generate different polyhedral feasible regions for investors' asset allocation decisions. Results showed that the data-driven problem led to less conservative solutions than classical RO.

To control the downside of losses, the  lower partial moment (LPM) can also be used, which is more perceivable by investors than other risk measures. \cite{ling2019robust} proposed a multi-period PSP similar to that of \cite{dantzig1993multi} based on a downside risk measure with an asymmetrically distributed uncertainty set. The objective function combines the expected terminal wealth of the portfolio with its LPM.
At each period \(s=0,\dots,S\), returns are denoted as \(r_{0}^{s},r_{1}^{s},...,r_{n}^{s}\), where \(r_{0}^{s}\) is the deterministic risk-free return and \(r_{j}^{s}\) is the uncertain return of risky asset $j$. The decision variable \(x_j^s,\, j=0,\dots,n\) denotes the dollar amount invested in asset \(j\) in period \(s\). With that, the terminal value of the portfolio is given by \(w^{S}=x_{0}^{S}(1+r_{0}^{S})+(e+r^{S})'x^{S}\), and the objective function is \(\min{-\mathbb{E}[W^{T}]+\lambda .\mathbb{E}[(\alpha - W^{T})_{+}]}\). Rebalancing constraints, similar to those used in \cite{dantzig1993multi}, are included.

Risk in a multi-period PSP can also be captured by the volatility of terminal wealth using mean-variance multi-period PSPs.
\cite{cong2017robust} considered discrete periods, indexed by \(s \in \{0,\Delta s,...,S-\Delta s\}\) for investment and denote by \(S\) the terminal period. Their formulation is based on maximizing the expected terminal wealth and minimizing the investment risk, quantified as \(\hat{v}_{0}(W_{0})=\max_{\{\hat{x}_{s}\}_{s=0}^{S-\Delta s}}\{E[W_{S}|W_{0}]-\lambda.Q[W_{s}|W_{0}]\}\), where \(\hat{v}\) is the value function. In this formulation, \(W\) is the wealth, which is calculated as \(W_{s+\Delta s}=W_{s}.(\hat{x}_{s}'r_{t}^{e}+r_{f}), \;s=0,\; \Delta s,...,S-\Delta s\), whereas \(r_{f}\) is the return of the risk-free asset and \(r_{s}^{e}=[r_{s}^{e}(1),...,r_{s}^{e}(n)]\) is the vector of returns of the risky assets during \([s,s+\Delta s]\). \cite{cong2017robust} argued that solving this problem using dynamic programming is difficult because of the non-linearity of conditional variance, so they replaced the dynamic mean-variance problem with a dynamic quadratic optimization problem. The new formulation is a target-based optimization since the risk aversion coefficient acts similar to an investment target in the problem. Moreover, solving the dynamic mean-variance PSP based on target-based optimization ensures time-inconsistency or ``pre-commitment strategy", which means that the investor has committed to an initial investment strategy. However, in many cases, investors do not want to commit to an initial investment strategy. Therefore, \cite{basak2010dynamic} suggested a time-consistency restriction in the formulation that can be solved in a backward recursive manner. Nevertheless, in both cases of pre-commitment and time-consistency strategies, the mean vector and the covariance matrix of returns of risky assets are subject to uncertainty. Hence, \cite{cong2017robust} proposed robust pre-commitment and time-consistency strategies where stationary and non-stationary formulations generate portfolios with the same Sharpe ratio given the risk-free asset as a benchmark. \cite{jiang2021robust} proposed a multi-period, multi-objective PSP where the objectives are the expected value and variance of the portfolio returns. To consider parameter uncertainty, an ellipsoidal uncertainty set is used for asset returns, leading to a SOCP. Moreover, a weighted-sum approach is used to obtain the Pareto frontier of the solutions.

Volatility measures can be used to define an arbitrage opportunity, which is a portfolio that can be formed with a negative investment while its profit is positive. \cite{pinar2005robust} considered \(n\) risky assets, where \(\nu_{j}\) is the period-end value of \(\$1\) invested in asset \(j\) at the beginning of the period. They used \(\nu=(\nu_{1},...,\nu_{n})\) as the vector of the end-of-period values, which \(\bar{\nu}\) is its expected value and \(Q\) is its covariance matrix. The vector of return is defiend as \(r=\nu-e\). If \(\nu\) is known in advance, a portfolio \(x\) that satisfies \(\bar{\nu}^{'}x\geq 0,\; x^{'}Qx=0,\; e^{'}x\leq 0\)
corresponds to an arbitrage opportunity. Since \(x'Qx=0\) then there is not any deviation in the return of assets form their expected values. These conditions mean that there is a portfolio that can be formed with a negative investment while its profit is positive. In practice \(x'Qx\) cannot be equal to zero. An investor can assume that a random number is "rarely" less than its mean minus \(\theta\) times of its standard deviation as \(\bar{\nu}^{'}x-\theta \sqrt{x^{'}Qx}\geq 0,\; e^{'}x\leq 0\). \cite{pinar2005robust} demonstrated that these conditions are related to an RO approach with an ellipsoidal uncertainty set. They also developed a multi-period PSP formulation by defining a self-financing constraint, in which the investment amount in the second period is based on the income of the first period. The end-of-period value of \(\$1\) invested in an asset at each period is uncertain and belong to an ellipsoidal uncertainty set. An adjustable RO approach was used to handle uncertainty. 

While most robust PSPs are modelled under the assumption that investors are perfectly rational beings, \cite{liu2015robust} argued that the rationality assumption does not always hold. Studies of behavioral finance have found that the axioms of rationality are violated across a range of financial decision-making situations. The prospect theory delineates the behavior of investors and asserts that investors value gains and losses differently. \cite{liu2015robust} proposed a robust multi-period PSP based on the premises of the prospect theory. Instead of classical utility or disutility functions, an S-shape value function, originally introduced by \cite{kahneman2013prospect}, is used to model the investor perception towards return. To account for uncertainty in cumulative asset returns, a budget uncertainty set whose level of conservatism can be controlled is utilized. However, applying the prospect theory value function leads to a complex nonlinear programming model that is intractable. Therefore, an improved particle swarm optimization (PSO) algorithm was used to solve the problem. 

Besides the uncertainty of individual asset returns at each period, macroeconomic conditions represent another source of uncertainty. 
\cite{desmettre2015robust} proposed a formulation for a multi-period investment problem under uncertainty introduced by uncertain market crash sizes in an interval. The objective is to maximize the terminal wealth. This problem uses a min-max worst-case scenario formulation that can be solved analytically. However, an interval uncertainty set results in over-conservative solutions. 

Another way to represent the uncertainty of parameters is by using discrete scenarios, which often lead to less complex formulations compared to those based on continuous uncertainty sets. The next section focuses on the use of discrete scenarios in robust multi-period PSPs.

\subsection{Robust Discrete Scenarios and Decision Tree Models}\label{sec:descrete}

\cite{mulvey1995robust} developed a robust framework based on discrete scenarios in which infeasibility is allowed under some scenarios but is penalized in the objective function. Application of this approach to robust multi-period PSPs usually leads to less complex formulations than for robust problems that use continuous uncertainty sets. \cite{pinar2007robust} considered a two-period PSP in which the returns of risky assets are uncertain, and used a discrete scenario tree to model uncertainty. In another attempt, \cite{oguzsoy2007robust} proposed a robust multi-period PSP with rebalancing and transaction costs. The problem is formulated as an MIP since its decision variables are the number of shares. They also developed a scenario-based, multi-period, mean-variance PSP, in which a decision tree with different levels is used. Portfolio rebalancing can happen at any level of the decision tree, and each tree node shows different rival scenarios for returns and risk (variance). A min-max formulation is used to find the worst-case robust solution. The robust counterpart considers risk scenarios at each node, time period and return realization . Since it is very unlikely that the worst scenario across all dimensions is realized, this approach leads to overly conservative solutions. Conversely,
\cite{shen2008robust} used semi-variance as a disutility function (\ie risk measure), which penalizes only negative deviations. Both asset returns in each scenario and the conditional probabilities of scenarios are treated as uncertain parameters. Ellipsoidal uncertainty sets are  for returns of assets at each scenario, which leads to a SOCP. 

Two-stage stochastic programming is a practical framework for modeling uncertainty in optimization problems. In this approach, decision variables are divided into ``here-and-now" and ``wait-and-see" variables. The mathematical formulation of a two-stage stochastic programming is \(\min_{x\in X}c'x+\mathbb{E}[F(x,\xi(w))]\), where \(F(x,\xi(w))=\min {f(w)'y},\;s.t.\;A(w)x+Dy=b(w),\; y\geq 0\), where \(x\) is a ``here and now" decision variable, \(y\) is a ``wait and see" decision variable, \(\mathbb{E}(.)\) is the expected value, \(\xi(w)=(f(w),A(w),b(w))\) is the uncertain vectors, and \(D\) is the fixed recourse matrix. \cite{ling2017robust} argued that because two-stage stochastic programming is a risk-neutral approach, it is not suitable for a certain setting, and developed a two-stage stochastic program with a mean-risk aversion concept as \(\min_{x\in X}c'x+\mathbb{E}[F(x,\xi(w))]+\lambda \rho(F(x,\xi(w)))\), where \(\rho\) is a risk measure and \(\lambda\geq 0\) is a trade-off coefficient that captures the risk-aversion attitude of the decision maker. To tackle the same problem, \cite{ahmed2006convexity} used variance as a risk measure, while \cite{ling2017robust} used CVaR as a risk measure leading to the less complex formulation \(\min_{x\in X}c'x+\mathbb{E}[F(x,\xi(w))]+\lambda CVaR_{\alpha}(F(x,\xi(w)))\). \cite{ling2017robust} assumed that asset returns in the first stage belong to a set of scenarios with known probabilities, whereas the distribution function of asset returns in the second stage belongs to an ambiguity set with uncertainty about the first two moments. This approach results in a SDP formulation. 
Even though using discrete scenarios and a decision tree for a multi-period PSPs lead to tractable formulations, identifying all possible scenarios might be challenging.  

\subsection{Robust Regime Dependent Models}

\cite{liu2014regime} argued that stock prices are affected by market conditions, which are assumed to follow a Markov regime-switching process. Specifically, in each market regime, financial parameters have different distribution functions. An approach to deal with parameter uncertainty in different market conditions is by using regime-dependent robust formulations. \cite{liu2014regime} described the time-varying properties of random returns by using a nonlinear dynamic model between periods. They assumed different uncertainty sets for each market situation. VaR is used as the basic risk measure in the formulation, where the distribution function of asset returns is uncertain and belongs to a moment-based ambiguity set. A restrictive assumption made in this study is that uncertainty sets of adjacent periods are independent and static, whereas in reality they usually are dynamic and dependent. 
\cite{liu2018time} considered dependency of dynamic uncertainty sets between adjacent periods in their formulation. Moreover, instead of VaR, they used CVaR as the risk measure. Similar to the formulation of \cite{liu2014regime}, it is assumed that moments of the loss function distribution are known and fixed, which leads to a SOCP formulation.

\cite{yu2016regime} also applied the regime-switching uncertainty set approach on a mean-CVaR PSP where the loss function is assumed to be the difference in wealth between times $s-1$ and $s$. This practically means that at each period, there is a different loss function which results in a different CVaR constraint. Because at each market state the risk-free rate of return can also change, risky asset returns, risk-free asset returns, and the distribution function of the loss function (probability of each scenario) are assumed to be uncertain and belong to ellipsoidal uncertainty sets. A three-step algorithm is used to find optimal solutions of the multi-period PSP based on different market states. An important advantage of this multi-period PSP is that it captures both regime-switching and parameter uncertainty simultaneously, leading to a more practical formulation than classical robust multi-period PSPs.

\subsection{Asset-Liability Management Problem}

Asset Liability Management (ALM) entails the allocation and management of assets, equity, interest rate, and credit risk (including risk overlays) to cover the commitments (\ie debts). In this section, we survey applications of RO in ALM problems.

\cite{van2007optimal} demonstrated that there are two types of investment strategies in an ALM problem: passive risk management, and active risk management. In the passive strategy, allocation of budget among different benchmarks such as equity, bonds, real estate \etc is the main decision. In active risk management, decisions are about tactical and operational investment activities that involve a number of investment managers, each is assigned a specific benchmark category. A formulation that calculates the total return of each manager by solving a mean-variance PSP based on the calculated expected value and variance of investment returns is proposed. These parameters are assumed to belong to ellipsoidal uncertainty sets. Practically, this robust ALM problem is a mean-variance PSP while the expected return and variance of asset returns belong to uncertainty sets.  

\cite{iyengar2010robust} assumed that the source of uncertainty of asset returns are fundamental factors. Then, a factor model can capture the true uncertainty of asset returns instead of predefined nominal asset returns. Using robust factor models in ALM problems can enable the true sources of uncertainty to be captured, leading to more realistic formulations with better out-of-sample performance. 
\cite{iyengar2010robust} developed a RO formulation for pension fund management, which is an ALM problem with a constraint on funding ratio. 
This ratio indicates the value of assets to the present value of liabilities that are used in a chance constraint, where the probability that funding ratio is greater than a threshold should be greater than a confidence level. The present value of liabilities depends on the interest rate, whereas asset values depend on their rate of return. In the proposed formulation, the funding ratio is assumed to be an uncertain parameter that follows a factor model by a function that defines stochastic parameters. A Gaussian process for factors of uncertain parameters is considered. Parameters of factor models are assumed to belong to an ellipsoidal uncertainty set, which results in a SOCP. \cite{platanakis2017asset} proposed a factor model for asset returns and liabilities in which factor loading belongs to an ellipsoidal uncertainty set, asset returns and liabilities belong to box uncertainty sets, and the covariance matrix of disturbances has upper and lower bounds on its elements. It has been shown that this problem can be reformulated into a SOCP. 

\cite{gulpinar2013robust} used time-varying investment opportunities to propose a robust ALM. This method assumes that a future rate of return of an asset depends on its rate of return in a former period. They augmented the multi-period PSP formulation of \cite{dantzig1993multi} by adding liabilities and a funding ratio constraints. The transformation of \cite{ben2000robust} was also used to simplify the formulation, by which the cumulative rates of return of assets are the uncertain parameters that belong to an ellipsoidal uncertainty set. Asset returns and interest rates are assumed to follow a vector-autoregressive (VAR) process that captures the time-varying aspect of investment. Unlike the symmetric uncertainty sets assumption in other robust ALM problems, \cite{gulpinar2016robust} developed a robust ALM problem using asymmetric uncertainty set, which captures the structure of uncertainty more accurately. Recently, \cite{gajek2021robust} proposed a robust ALM formulation where the interest rate is uncertain and the distribution function of the uncertain parameters belongs to a nonempty ambiguity set. This formulation bounds from above VaR of the change in the portfolio value due to interest rate model violation.

\subsection{Robust Control Formulation}

Robust control methods are designed to function properly provided that the uncertain parameters or disturbances are contained within some bounded/compact sets. \cite{flor2014robust} developed a robust control formulation for an investment PSP. They assumed that an investor has access to stocks, bonds, and cash while interest rates are uncertain. In this formulation, a robust control, time-continuous formulation for the uncertainty of interest rate is developed. Results showed that the proposed model is more sensitive to the ambiguity about stocks than bonds. This problem is time-continuous, thus is formulated using differential equations. 

\cite{glasserman2013robust} developed a robust control formulation for a multi-period PSP based on a factor model that is used to calculate the return of assets at the next period. They assumed that the factors are mean reverting and evolving and that their value at any time is a function of their previous time value and its residual. Two regression models were used. In the first model, a factor model calculates the return of assets at the next period. The second factor model calculates factor values at the next period. Sources of uncertainty in this PSP are the residuals of the two factor-models. Based on this formulation, the goal is to maximize the net present value of risk-adjusted excess gains by considering restricted transaction costs. Moreover, models are developed in two cases; finite-horizon investment, and infinite-horizon investment. A robust formulation based on the Bellman equation, leading to a dynamic programming model, is used. Results showed that the robust control formulation of \cite{glasserman2013robust} is more robust than deterministic formulations against perturbations of uncertain parameters. 

\cite{bo2017robust} applied a robust control approach for the credit portfolio, where the impact of credit risk model misspecification on the optimal investment strategies is measured. They proposed a formulation for a dynamic credit portfolio that accounts for robust decision rules against misspecifications of a model for the actual default intensity. Default intensity is defined as the probability of default for a certain time period conditional on no earlier default. In this formulation, an investor can invest in the money market and bonds by a pricing model of bonds that considers credit intensity. This portfolio formulation tries to maximize wealth while default intensity is uncertain.

\section{Other Financial Problems}\label{sec:special}

In this section, special PSP formulations are reviewed, including Log-robust portfolio selection, robust index tracking, hedging formulation, risk-adjusted Sharpe ratio, scenario-based formulation, and robust data envelopment analysis (DEA) for PSPs.

\subsection{Log-Robust Portfolio Selection} \label{4}

\cite{hull2003options} defined the Log-return as the equivalent, continuously-compounded rate of return of asset returns over a period of time. Log-return is calculated by taking the natural log of the ending stock price divided by the beginning value. It is based on a Levy process that represents the movements of a stock price whose successive displacements are random, independent, and statistically identical over different time intervals of the same length. Assume that Log-return of stock \(j\) at time \(S\) can be described as \(Ln\frac{Pr_{j}(S)}{Pr_{j}(0)}=(\mu_{j}-\frac{\sigma_{j}^{2}}{2})S+\sigma_{j}\sqrt{S}z_{j}\),
where \(S\) is the length of the time horizon, \(Pr_{j}(0)\) is the initial price of stock \(j\), \(Pr_{j}(S)\) is the stock price at time \(S\), \(\mu_{j}\) is the drift of the Levy process for stock \(j\), and \(\sigma_{j}\) is the standard deviation of the Levy process for stock \(j\). \cite{kawas2011log} proposed a Log-robust PSP where the scaled deviation belongs to a budget uncertainty set in two cases: correlated and uncorrelated assets. Let the uncertainty be represented as \(\sum_{j=1}^{n}|\tilde{z}_{j}|\leq \Gamma,\;|\tilde{z}_{j}|\leq 1,\;\forall j\). Then, the robust problem can be formulated as \(\max_{\tilde{x}} \min_{\tilde{z}} \sum_{j=1}^{n}\tilde{x}_{j}Pr_{j}(0)exp[(\mu_{j}-\frac{\sigma_{j}^{2}}{2})S+\sigma_{j}\sqrt{S}x\tilde{z}_{j}],\;\text{s.t.}\;\sum_{j=1}^{n}|\tilde{z_{j}}|\leq \Gamma,\;|\tilde{z_{j}}|\leq 1\;\forall j,\;\sum_{j=1}^{n}\tilde{x}_{j}Pr_{j}(0)=B_{0},\;\tilde{x}_{j}\geq 0\; \forall j\), where \(B_{0}\) is available budget. \cite{kawas2011log} transformed this formulation into an LP. They also considered a PSP with correlated assets, where \(Ln\frac{Pr_{j}(S)}{Pr_{j}(0)}=(\mu_{j}-\frac{\sigma_{j}^{2}}{2})T+\sqrt{S}Z_{j}\), where \(Z\) has normal distribution with mean \(0\) and covariance matrix \(Q\). They defined \(Y=Q^{\frac{-1}{2}}Z\), where \(Y\sim N(0,I)\). \cite{kawas2011log} proposed a tracktable robust counterpart in the case correlated assets. \cite{kawas2011short} extended the Log-robust PSP by allowing short selling, whereas \cite{pae2014log} added a transaction cost constraint to make the formulation more realistic. Instead of using predefined uncertainty sets, \cite{kawas2017log} proposed a data-driven Log-robust PSPs for two cases, correlated and uncorrelated assets. In both cases, they optimized the worst-case PSP over the worst of finitely many polyhedral uncertainties sets using different estimation methods. Consequently, both the uncertainty of parameters and the ambiguity of uncertainty sets are considered. However, the robust formulations are based on the worst-case perspective and
the solutions are still over-conservative. In contrast, \cite{lim2012robust} proposed a relative robust log-return PSP which is less conservative than the worst-case Log-robust PSP, yet harder to solve.

\cite{gulpiotanar2014robust} studied the robust PSP under supply disruption in the petroleum markets based on Log-return. They proposed a framework for portfolio management with a combination of commodities and stocks when the supply of commodities is uncertain. A geometric mean-reverting jump process is considered for prices to model the jumps (\ie large discrete movements). Both symmetric (ellipsoidal, and $D$-norm uncertainty sets) and asymmetric uncertainty sets for uncertain parameters are used. Results show that the $D$-norm uncertainty set leads to more extreme portfolio allocations with less diversification than the ellipsoidal and asymmetric uncertainty sets. Moreover, the asymmetric uncertainty set with a high price of robustness results in a high level of diversification.

\subsection{Index-Tracking Portfolio Selection} \label{5}

Index tracking is a passive investment strategy where a portfolio is formed to follow an index benchmark. Hence, a logical index-tracking portfolio includes all stocks under an index based on their value weights. However, the need for frequent re-balancing transactions to closely track the index might lead to high transaction costs. Therefore, decision-makers might try to find the best possible combination of assets that follows a benchmark index with the lowest possible transaction cost, while also accounting for parameter uncertainty. \cite{costa2002robust} developed two robust index-tracking PSPs where the return vector and the covariance matrix of risky assets are uncertain and belong to polytope uncertainty sets. Practically, the variance of tracking error (\ie the difference in actual performance between a portfolio and its corresponding benchmark) is used to capture the volatility of tracking error, leading to a quadratic programming (QP) formulation. In this formulation, the return of a given portfolio \(x\) is calculated as \(x'r+(1-x')r_{f}\), whereas the return of the benchmark portfolio (index), denoted by \(x_{B}\), is calculated as \(x_{B}'r+(1-x_{B}')r_{f}\). With that, the tracking error is calculated as \(tr(x)=(x-x_{B})'r+(x_{B}-x)'r_{f}\), and the expected value and the variance of tracking error are \(\rho_{\varphi}(x)=(x-x_{B})'r+(x_{B}-x)'r_{f}=(x-x_{B})'(\hat{r}-r_{f})\), and \(\sigma_{Q}^{2}(x)=(x-x_{B})'Q(x-x_{B})\), respectively. Hence, the problem is formulated as \(\min_{x\in X}\;\sigma_{Q}^{2}(x)\;\text{s.t.}\;\rho_{\varphi}(x)\geq E\), where \(E\) is the minimum acceptable target for the expected value of tracking error. \cite{costa2002robust} assumed that \(r\), \(r_{f}\), and \(Q\) are not exactly known. Thus, they defined a set of all possible matrices \(\Phi\in Con{[\Phi_1,...,\Phi_n]}\) where \(\Phi=\begin{pmatrix}
Q & r \\
0 & r_{f} 
\end{pmatrix}\) and showed that the robust index tracking formulation can be transformed to a tractable formulation by using a linear matrix inequality. However, estimating the covariance matrix is computationally expensive in large problems. Hence, instead of the variance of tracking error, \cite{chen2012robust} proposed a robust similarity measure that measures pairwise similarities between the assets and the targeted index, with a budget uncertainty set. Moreover, a cardinality constraint is used to limit the number of assets in the optimal portfolio, leading to a MIP.

The aforementioned robust index-tracking PSPs ignore the distribution function of asset returns. Alternatively, one can use partial information about the distribution of asset returns based on historical data. \cite{ling2014robust} developed a distributionally robust downside risk measure formulation for index-tracking PSPs with a moment-based ambiguity set in two cases: 1) the first two moments (mean and covariance) are known and fixed, 2) the first two moments belong to ellipsoidal and polyhedral uncertainty sets, respectively. Results demonstrate that the distributionally robust index-tracking PSP provides less conservative solutions than classical robust index-tracking PSPs.

\subsection{Robust Hedging}

Hedging means an investment position intended to offset potential losses or gains that may be incurred by a companion investment. Options are important financial tools used for Hedging risk. An option is the opportunity, but not the obligation, for buying or selling underlying assets. \cite{lutgens2006robust} used options to propose a RO formulation for hedging risk in two cases: a single stock and an option, and multiple assets and options. In the former case, they optimized the expected return while assuming that asset returns belongs to a discrete (scenario-based) uncertainty set. This formulation led to a max-min problem with a nonlinear inner optimization problem. In the second case, they assumed that the return vector belongs to an \(N\)-dimensional ellipsoidal uncertainty set, which results in a SOCP.

\cite{gulpinar2017robust} used weather derivatives in a PSP in which CVaR is the risk measure. 
Weather derivatives are traded as financial instruments between two parties. The seller agrees to bear the risk for a premium and makes a profit if nothing happens. However, if the weather turns out to be bad, then the buyer claims the agreed amount. The price of this specific derivative is a function of the weather. \cite{gulpinar2017robust} suggested a spatial temperature modeling where the correlation between the locations of weather derivatives under consideration are explicitly taken into account. Both symmetric (ellipsoidal) and asymmetric uncertainty sets are used to develop robust counterparts. Experimental results showed that a robust model with weather derivatives has better performance in the worst-case analysis.

\subsection{Robust Sharpe and Omega Ratio}

The Sharpe ratio (SR) is defined as a ratio of the expected excess return over the risk-free rate to the standard deviation of the excess return. However, parameters of the Sharpe ratio are subject to uncertainty. In practice, an estimate of the Sharpe ratio is used in optimization problems. To mitigate the estimations error, \cite{deng2013robust} proposed a robust risk-adjusted Sharpe ratio and a robust VaR-adjusted Sharpe ratio (VaRSR), defined as the lowest Sharpe ratio consistent with the data in the observation period for a given confidence level. Based on the normality assumption of asset returns, \cite{zymler2011robust} argued that an uncertainty set for a Sharpe ratio can take the form of an ellipse with exogenous parameters. They then showed that in one dimension, the uncertainty set is an interval where the inner-optimization solution in the robust formulation of a Sharpe ratio is exactly equal to a risk-adjusted Sharpe ratio. Results showed that VaRSR is more robust than SR when the return distribution is non-normal. 

Maximizing Sharpe ratio is an important performance measures in PSPs. However, PSPs are prone to estimation errors and optimization amplifies estimation errors, resulting in portfolios with poor out-of-sample performance. One way to deal with this drawback is combination portfolios. Here, the portfolio is a linear combination of two or more prespecified portfolios. A proper combination can improve Sharpe ratio of the portfolio. \cite{chakrabarti2021parameter} proposed a combination of robust minimum-variance and maximum Sharpe ratio based on a robust regret-minimizing portfolio. They used box uncertainty sets for the asset returns and the covariance matrix. Finally, each portfolio is scored based on its worst-case regret and the optimal portfolio is the one with the smallest worst-case regret. Results showed that this portfolio is relatively close to the optimal combination portfolio for the actual parameter values.

Omega, an important ratio in finance proposed by \cite{keating2002introduction}, is the ratio of risk to return, assuming there is a predetermined threshold that partitions the returns into losses and gains.
This ratio is an alternative to the Sharpe ratio and is based on information the Sharpe ratio discards. In practice, Sharpe ratio considers only the first two moments of the return distribution while Omega ratio considers all moments. \cite{kapsos2014optimizing} showed that the Omega ratio can be represented into an LP. \cite{kapsos2014worst} introduced the worst-case Omega ratio (WCOR) when distribution functions of asset returns are partially known and belong to three different ambiguity sets. First, the underlying distribution is a mixture distribution with known continuous mixture components but unknown mixture weights. The second ambiguity set encompasses all possible distributions supported on a discrete set of scenarios. The third one uses box and ellipsoidal uncertainty sets for the probabilities of scenarios. Even though the Omega ratio considers both losses and gains,
\cite{sharma2017omega} argued that this approach is too sensitive to threshold used. Moreover, there is not any systematic way to specify this threshold. The formulation in \cite{ghahtarani2019mathematical} uses the fundamental value of an asset as a threshold of the Omega ratio, which protects the portfolio against bubble conditions in the market. \cite{sharma2017omega} redefined the Omega ratio by using a loss function instead of the return. Hence, it minimizes losses greater than a threshold and maximizes losses less than the same threshold when CVaR is used as the threshold. Furthermore, they developed a distributionally robust Omega-CVaR optimization formulation in which the probability of each scenario of the loss function is uncertain and belongs to three uncertainty sets: a mixed uncertainty set, a box uncertainty set, and an ellipsoidal uncertainty set. The first two uncertainty sets lead to LPs, whereas the last one results in a SOCP. \cite{yu2019realized} compared results of the worst-case Omega ratio to those of the worst-case CVaR and relative CVaR formulations while adding transaction costs constraint and allowing short selling. Results show that the worst-case Omega portfolio yields lower loss values and higher market values compared to CVaR-based models under various confidence levels. \cite{georgantas2021robust} also compared the robust Omega ratio PSP proposed by \cite{kapsos2014worst} to the robust mean-variance PSP with box and ellipsoidal uncertainty sets and the robust CVaR PSP proposed by \cite{zhu2009worst}. Results showed that robust PSPs are less diversified than their nominal counterparts. However, improvements were observed in the portfolio performance. Another comparison in this context has been done by \cite{sehgal2021robust}. They compared PSPs based on robust Omega ratio, semi-mean absolute deviation ratio, and weighted stable tail adjusted return ratio (STARR) with their non-robust counterparts. In these formulations, a budgeted uncertainty set is used for asset returns. Results showed that the robust formulations outperform the nominal problems with respect to standard deviation, value at risk (VaR), conditional value at risk (CVaR), Sharpe ratio, and stable tail adjusted return ratio (STARR).

Sharpe and Omega ratios are based on the absolute volatility of assets. However, some investors make decisions based on the volatility of an asset compared to the market and not on the absolute volatility itself. Beta is a measure of volatility that indicates whether an asset is more or less volatile compared to the market. Hence, Beta can be used as a decision criteria to capture the volatility of an asset compared to the market. A asset's Beta is calculated by dividing the product of the covariance of the asset returns and the market returns by the variance of the market returns over a specified period. However, this measure is subject to uncertainty since all components of the Beta formula are uncertain parameters. \cite{ghahtarani2013robust} proposed a robust multi-objective PSP where the objectives are the portfolio rate of return and its systematic risk (Beta). A budget uncertainty set to model the uncertainty of Beta is used. The problem is reformulated as a tractable goal program. Results show that portfolios selected based on the robust Beta outperform non-robust Beta portfolios in terms of weight stability and return volatility.

\subsection{Robust Scenario-Based Formulation}

Unlike Section 5.2 that reviews scenario-based formulations of multi-period PSPs, this section focuses on the use of discrete scenarios to represent uncertainty in single-period PSPs.
\cite{kouvelis1997robust} proposed a robust formulation for a discrete scenario-based uncertainty set. It optimizes an objective function based on the worst possible scenario, which leads to the worst-case conservative results. \cite{roy2010robustness} proposed a new definition for robust scenario-based solutions in which a solution is robust if it exhibits good performance in most scenarios without ever exhibiting very poor performance in any scenario. Then, they developed \emph{\(bw\)-robustness}, by taking into consideration minimum acceptable objective value and a target objective value to achieve, or exceed if possible. \cite{gabrel2018portfolio} developed a robust scenario-based PSP by using both worst-case scenario and \(bw\)-robustness to maximize a portfolio's return while returns of assets belong to a discrete scenario-based uncertainty set. Moreover, they introduced a new robustness criterion called \emph{\(pw\)-robustness}, in which instead of maximizing a proportion of scenarios that their values are greater than or equal to a threshold, the decision-maker specifies a fixed proportion of scenarios, and maximizes the value of the soft bound.
The \(pw\)-robustness formulation is a MIP. To circumvent the computational time issue, \cite{gabrel2018portfolio} proposed two heuristic methods that can be used to obtain quick solutions for problems of large sizes. 

Some investors might invest based on their preferences of assets, where ranking information of assets are uncertain. \cite{nguyen2012robust} proposed a robust ranking mean-variance, which is similar to the classical mean-variance. However, the ranking of assets is used instead of the return of assets. Formulations were developed in two cases: the maximum ranking with and without risk (variance). The ranking of assets belongs to a discrete uncertainty set, which leads to a MIP solved by a constraint generation method. 

\subsection{Robust Data Envelopment Analysis and Portfolio Selection}

One way for evaluating stocks or assets in the financial markets is data envelopment analysis (DEA), in which the efficiency of stocks or assets is evaluated based on a set of inputs and outputs (criteria). Based on this method, units (assets or stocks) are divided into two parts: efficient, and inefficient. Consequently, DEA calculates the efficacy rate of units. \cite{peykani2016utilizing} demonstrated that the efficiency of stocks in DEA depends on inputs and outputs, which are uncertain. Consequently, they proposed robust DEA with a budget uncertainty set. Results of robust DEA are more robust than a non-robust DEA formulation with respect to the efficiency of stocks. However, their formulations can be used only for continuous uncertainty sets.
\cite{peykani2019stock} developed a robust DEA for a discrete scenario formulation with uncertainty, which expands the application of robust DEA to financial problems in the real-world. However, these robust DEA formulations provide the efficiency ratio without any detail about the amount of money invested in each asset while an investor needs to know the proportion of investment of funds invested in each asset. \cite{peykani2020novel} proposed a two-phase portfolio selection process. At the first stage, the efficiency of candidate stocks is evaluated by robust DEA. In the second stage, the optimal portfolio is formed by using robust mean-semi variance-liquidity and robust mean-absolute deviation-liquidity models. In both phases, budget uncertainty sets are used for the uncertain parameters. This two-phase formulation provides two filters (robust DEA, and robust PSP) to find the optimal portfolio.

Table \ref{table:50} lists all the reviewed articles (n=142) in a chronological order and classifies them based on the problem type (PSP), uncertain parameters (UP), the structure of uncertainty or ambiguity sets used in the robust formulation (U/A set), the robust optimization method employed to deal with uncertainty (RO method) and the class of the tractably reformulated problem (Model).

%\begin{landscape}
\begin{center}
\footnotesize{
\begin{longtable}{l p{4cm} ccccccccccccc cccccc cccccc ccccc ccccccc}
\label{table:50}
\\ \caption{Summary of the reviewed articles}
   \\ \hline
    \multicolumn{1}{c}{\begin{turn}{-90}\(\#\)\end{turn}}&\multicolumn{1}{c}{ Article}&\multicolumn{13}{c}{PSP type}&\multicolumn{6}{c}{UP \footnote{Uncertain Parameters}}&\multicolumn{6}{c}{U/A \footnote{Uncertainty or Ambiguity} set}&\multicolumn{5}{c}{RO Methods}&\multicolumn{7}{c}{Model}\\
    \cline{3-39}
    & &\begin{turn}{-90}Mean-Variance\end{turn}&\begin{turn}{-90}Minimum Variance\end{turn}&\begin{turn}{-90}MAD\end{turn}&\begin{turn}{-90}LPM\end{turn}&\begin{turn}{-90}Factor-Based\end{turn}&\begin{turn}{-90}Utility function\end{turn}&\begin{turn}{-90}VaR/CVaR\end{turn}&\begin{turn}{-90}Multi-Period\end{turn}&\begin{turn}{-90}ALM\end{turn}&\begin{turn}{-90}Log-return\end{turn}&\begin{turn}{-90}Index tracking\end{turn}&\begin{turn}{-90}Ratios\end{turn}&\begin{turn}{-90}Others\end{turn}&\begin{turn}{-90}Asset return\end{turn}&\begin{turn}{-90}Variance-Covariance matrix\end{turn}&\begin{turn}{-90}Factor-based parameters\end{turn}&\begin{turn}{-90}Distribution function\end{turn}&\begin{turn}{-90}Scale parameter (Log-Return)\end{turn}&\begin{turn}{-90}Others\end{turn}&\begin{turn}{-90}Classical\end{turn}&\begin{turn}{-90}Asymmetrical\end{turn}&\begin{turn}{-90}Moment-based\end{turn}&\begin{turn}{-90}Metric-based\end{turn}&\begin{turn}{-90}Discrete \end{turn}&\begin{turn}{-90}Others\end{turn}&\begin{turn}{-90}RO\end{turn}&\begin{turn}{-90}DRO\end{turn}&\begin{turn}{-90}Relative robust\end{turn}&\begin{turn}{-90}Adaptive robust\end{turn}&\begin{turn}{-90}Others\end{turn}&\begin{turn}{-90}LP\end{turn}&\begin{turn}{-90}NLP\end{turn}&\begin{turn}{-90}SOCP\end{turn}&\begin{turn}{-90}SDP\end{turn}&\begin{turn}{-90}Mixed integer\end{turn}&\begin{turn}{-90}Other Convex\end{turn}&\begin{turn}{-90}Non-convex\end{turn}\\
    \hline\hline
    \endhead 
   % \bottomrule%
\hline%    
\multicolumn{39}{l}{Continued on Next Page}\\%
\hline%
\endfoot%
\hline%
\multicolumn{39}{l}{}\\%
%\hline%
\endlastfoot%
% \midrule
1	&	\cite{ben2000robust}	&		&		&		&		&		&		&		&	*	&		&		&		&		&		&	*	&		&		&		&		&		&	*	&		&		&		&		&		&	*	&		&		&		&		&		&		&	*	&		&		&		&		\\
\hline
2	&	\cite{costa2002robust}	&		&		&		&		&		&		&		&		&		&		&	*	&		&		&	*	&	*	&		&		&		&		&	*	&		&		&		&		&		&	*	&		&		&		&		&		&	*	&		&		&		&		&		\\
\hline
3	&	\cite{lauprete2003robust}	&	*	&		&		&		&		&		&		&		&		&		&		&		&		&	*	&		&		&		&		&		&		&		&		&		&		&		&		&		&		&		&	*	&		&	*	&		&		&		&		&		\\
\hline
4	&	\cite{goldfarb2003robust}	&	*	&		&		&		&	*	&		&	*	&		&		&		&		&	*	&		&		&		&	*	&		&		&		&	*	&		&		&		&		&		&	*	&		&		&		&		&		&		&	*	&		&		&		&		\\
\hline
5	&	\cite{ghaoui2003worst}	&		&		&		&		&	*	&		&	*	&		&		&		&		&		&		&		&		&		&	*	&		&		&		&		&	*	&	*	&		&		&		&	*	&		&		&		&		&		&		&		&		&		&		\\
\hline
6	&	\cite{tutuncu2004robust}	&	*	&		&		&		&		&		&		&		&		&		&		&		&		&	*	&	*	&		&		&		&		&	*	&		&		&		&		&		&	*	&		&		&		&		&		&	*	&		&		&		&		&		\\
\hline
7	&	\cite{pinar2005robust}	&		&		&		&		&		&		&		&	*	&		&		&		&		&		&	*	&		&		&		&		&		&	*	&		&		&		&		&		&	*	&		&		&	*	&		&		&		&	*	&		&		&		&	*	\\
\hline
8	&	\cite{ceria2006incorporating}	&	*	&		&		&		&		&		&		&		&		&		&		&		&		&	*	&	*	&		&		&		&		&	*	&		&		&		&		&		&		&		&		&		&	*	&		&		&	*	&		&		&		&		\\
\hline
9	&	\cite{lu2006new}	&		&		&		&		&	*	&		&		&		&		&		&		&		&		&		&		&	*	&		&		&		&	*	&		&		&		&		&		&	*	&		&		&		&		&		&		&		&		&		&	*	&		\\
\hline
10	&	\cite{pachamanova2006handling}	&		&		&		&		&		&		&	*	&		&		&		&		&		&	*	&	*	&		&		&		&		&		&	*	&		&		&		&		&		&	*	&		&		&		&		&		&		&	*	&		&		&		&		\\
\hline
11	&	\cite{lutgens2006robust}	&		&		&		&		&		&		&		&		&		&		&		&		&	*	&	*	&		&		&		&		&		&	*	&		&		&		&	*	&		&	*	&		&		&		&		&		&	*	&	*	&		&		&		&		\\
\hline
12	&	\cite{fabozzi2007robust}	&	*	&		&		&		&		&		&		&		&		&		&		&		&		&	*	&		&		&		&		&		&	*	&		&		&		&		&		&	*	&		&		&		&		&		&		&	*	&		&		&		&		\\
\hline
13	&	\cite{bienstock2007histogram}	&	*	&		&		&		&		&		&		&		&		&		&		&		&		&	*	&		&		&		&		&		&		&		&		&		&		&	*	&	*	&		&		&		&		&		&		&		&		&	*	&		&		\\
\hline
14	&	\cite{garlappi2007portfolio}	&	*	&		&		&		&		&		&		&		&		&		&		&		&		&	*	&	*	&		&		&		&		&		&		&		&		&		&		&		&		&		&		&	*	&		&	*	&		&		&		&		&		\\
\hline
15	&	\cite{calafiore2007ambiguous}	&	*	&		&	*	&		&		&		&		&		&		&		&		&		&		&		&		&		&	*	&		&		&		&		&		&	*	&		&		&		&	*	&		&		&		&		&		&		&		&		&	*	&		\\
\hline
16	&	\cite{popescu2007robust}	&		&		&		&		&	*	&		&	*	&		&		&		&		&		&		&		&		&		&	*	&		&		&		&		&	*	&		&		&		&		&	*	&		&		&		&		&	*	&		&		&		&		&		\\
\hline
17	&	\cite {huang2007robust}	&		&		&		&		&		&		&	*	&		&		&		&		&		&		&		&		&		&	*	&		&	*	&		&		&	*	&		&		&		&		&	*	&		&		&		&		&		&		&	*	&		&		&		\\
\hline
18	&	\cite{pinar2007robust}	&		&		&		&		&		&		&		&	*	&		&		&		&		&		&	*	&		&		&		&		&		&		&		&		&		&	*	&		&	*	&		&		&		&		&	*	&		&		&		&		&		&		\\
\hline
19	&	\cite{oguzsoy2007robust}	&	*	&		&		&		&		&		&		&	*	&		&		&		&		&		&	*	&	*	&		&		&		&		&		&		&		&		&	*	&		&	*	&		&		&		&		&		&		&		&		&	*	&		&		\\
\hline
20	&	\cite{van2007optimal}	&	*	&		&		&		&		&		&		&		&	*	&		&		&		&		&	*	&		&		&		&		&		&	*	&		&		&		&		&		&	*	&		&		&		&		&		&		&	*	&		&		&		&		\\
\hline
21	&	\cite{ma2008robust}	&		&		&		&		&	*	&		&		&		&		&		&		&		&		&		&		&	*	&		&		&		&	*	&		&		&		&		&		&	*	&		&		&		&		&		&	*	&		&		&		&		&		\\
\hline
22	&	\cite{natarajan2008incorporating}	&		&		&		&		&		&		&	*	&		&		&		&		&		&		&		&		&	*	&		&		&		&		&	*	&		&		&		&		&	*	&		&		&		&		&		&		&	*	&		&		&		&		\\
\hline
23	&	\cite{huang2008portfolio}	&		&		&		&		&		&		&	*	&		&		&		&		&		&		&		&		&		&	*	&		&	*	&		&		&		&		&		&	*	&		&	*	&		&		&		&	*	&		&		&		&		&		&		\\
\hline
24	&	\cite{quaranta2008robust}	&		&		&		&		&		&		&	*	&		&		&		&		&		&		&	*	&		&		&		&		&		&	*	&		&		&		&		&		&	*	&		&		&		&		&	*	&		&		&		&		&		&		\\
\hline
25	&	\cite{bertsimas2008robust}	&		&		&		&		&		&		&		&	*	&		&		&		&		&		&	*	&		&		&		&		&		&	*	&		&		&		&		&		&	*	&		&		&		&		&	*	&		&		&		&		&		&		\\
\hline
26	&	\cite{shen2008robust}	&		&		&		&		&		&		&		&	*	&		&		&		&		&	*	&	*	&		&		&		&		&		&	*	&		&		&		&	*	&		&	*	&		&		&		&		&		&		&	*	&		&		&		&		\\
\hline
27	&	\cite{chen2009robust}	&	*	&		&		&		&		&		&		&		&		&		&		&		&		&	*	&	*	&		&		&		&		&		&	*	&		&		&		&		&	*	&		&		&		&		&		&	*	&		&		&		&		&		\\
\hline
28	&	\cite{demiguel2009portfolio}	&	*	&		&		&		&		&		&		&		&		&		&		&		&		&	*	&		&		&		&		&		&		&		&		&		&		&		&		&		&		&		&	*	&		&	*	&		&		&		&		&		\\
\hline
29	&	\cite{schottle2009robustness}	&	*	&		&		&		&		&		&		&		&		&		&		&		&		&	*	&	*	&		&		&		&		&	*	&		&		&		&		&		&	*	&		&		&		&		&		&		&	*	&		&		&		&		\\
\hline
30	&	\cite{zhu2009worst}	&		&		&		&		&		&		&	*	&		&		&		&		&		&		&		&		&		&	*	&		&		&	*	&		&		&		&	*	&		&		&	*	&		&		&		&	*	&		&	*	&		&		&		&		\\
\hline
31	&	\cite{natarajan2009constructing}	&	*	&		&		&		&		&		&	*	&		&		&		&		&		&		&	*	&		&		&		&		&		&	*	&		&		&		&		&		&	*	&		&		&		&		&	*	&	*	&		&		&		&		&		\\
\hline
32	&	\cite{lutgens2010robust}	&		&		&		&		&	*	&		&		&		&		&		&		&		&		&	*	&		&		&		&		&		&		&		&		&		&		&	*	&	*	&		&		&		&		&		&	*	&		&		&		&		&		\\
\hline
33	&	\cite{natarajan2010tractable}	&		&		&		&		&		&	*	&		&		&		&		&		&		&		&	*	&	*	&		&	*	&		&		&	*	&		&	*	&		&		&		&	*	&	*	&		&		&		&		&		&	*	&		&		&		&	*	\\
\hline
34	&	\cite{huang2010portfolio}	&		&		&		&		&		&		&	*	&		&		&		&		&		&		&		&		&		&	*	&		&		&		&		&		&		&	*	&		&		&		&	*	&		&		&	*	&		&	*	&		&		&		&		\\
\hline
35	&	\cite{iyengar2010robust}	&		&		&		&		&	*	&		&		&		&	*	&		&		&		&		&		&		&	*	&		&		&		&	*	&		&		&		&		&		&	*	&		&		&		&		&		&	*	&		&		&		&		&		\\
\hline
36	&	\cite{iyengar2010robust}	&		&		&		&		&		&		&		&		&	*	&		&		&		&		&	*	&		&		&		&		&		&	*	&		&		&		&		&		&	*	&		&		&		&		&		&		&	*	&		&		&		&		\\
\hline
37	&	\cite{zymler2011robust}	&	*	&		&		&		&		&		&		&		&		&		&		&		&		&	*	&		&		&		&		&		&	*	&		&		&		&		&		&	*	&		&		&		&		&		&		&	*	&		&		&		&		\\
\hline
38	&	\cite{gregory2011robust}	&	*	&		&		&		&		&		&		&		&		&		&		&		&		&	*	&		&		&		&		&		&	*	&		&		&		&		&		&	*	&		&		&		&		&		&	*	&		&		&		&		&		\\
\hline
39	&	\cite{gulpinar2011robust}	&	*	&		&		&		&		&		&		&		&		&		&		&		&		&	*	&		&		&		&		&		&	*	&		&		&		&		&		&	*	&		&		&		&		&		&		&	*	&		&		&		&		\\
\hline
40	&	\cite{moon2011robust}	&		&		&	*	&		&		&		&		&		&		&		&		&		&		&	*	&		&		&		&		&		&	*	&		&		&		&		&		&	*	&		&		&		&		&	*	&		&		&		&		&		&		\\
\hline
41	&	\cite{chen2011tight}	&		&		&		&	*	&		&		&		&		&		&		&		&		&		&	*	&		&		&		&		&		&		&		&	*	&		&		&		&		&	*	&		&		&		&		&		&	*	&		&		&		&		\\
\hline
42	&	\cite{lu2011robust}	&		&		&		&		&	*	&		&		&		&		&		&		&		&		&		&		&	*	&		&		&		&	*	&		&		&		&		&		&	*	&		&		&		&		&		&		&	*	&		&		&		&		\\
\hline
43	&	\cite{chen2011worst}	&		&		&		&		&		&		&	*	&		&		&		&		&		&		&	*	&	*	&		&		&		&		&		&	*	&		&		&		&		&	*	&		&		&		&		&		&		&	*	&		&		&		&		\\
\hline
44	&	\cite{hellmich2011efficient}	&		&		&		&		&		&		&	*	&		&		&		&		&		&		&		&		&		&	*	&		&		&	*	&		&		&		&		&		&		&		&		&		&		&		&	*	&		&		&		&		&		\\
\hline
45	&	\cite{guastaroba2011investigating}	&		&		&		&		&		&		&	*	&		&		&		&		&		&		&	*	&		&		&		&		&		&	*	&		&		&		&		&		&	*	&		&		&		&		&	*	&		&	*	&		&		&		&		\\
\hline
46	&	\cite{kawas2011log}	&		&		&		&		&		&		&		&		&		&	*	&		&		&		&		&		&		&		&	*	&		&	*	&		&		&		&		&		&	*	&		&		&		&		&	*	&		&		&		&		&		&		\\
\hline
47	&	\cite{kawas2011short}	&		&		&		&		&		&		&		&		&		&	*	&		&		&		&		&		&		&		&	*	&		&	*	&		&		&		&		&		&	*	&		&		&		&		&	*	&		&		&		&		&		&		\\
\hline
48	&	\cite{zymler2011robust}	&		&		&		&		&		&		&		&		&		&		&		&	*	&	*	&	*	&		&		&		&		&		&	*	&		&		&		&		&		&	*	&		&		&		&		&		&		&	*	&		&		&		&		\\
\hline
49	&	\cite{fonseca2012robust}	&	*	&		&		&		&		&		&		&		&		&		&		&		&		&	*	&		&		&		&		&	*	&	*	&		&		&		&		&		&	*	&		&		&		&		&		&		&	*	&		&		&		&		\\
\hline
50	&	\cite{fonseca2012robusta}	&	*	&		&		&		&		&		&		&		&		&		&		&		&		&	*	&		&		&		&		&	*	&	*	&		&		&		&		&		&	*	&		&		&		&		&		&		&	*	&		&		&		&		\\
\hline
51	&	\cite{sadjadi2012robust}	&	*	&		&		&		&		&		&		&		&		&		&		&		&		&	*	&		&		&		&		&		&	*	&		&		&		&		&		&	*	&		&		&		&		&		&		&		&		&	*	&		&		\\
\hline
52	&	\cite{garcia2012time}	&	*	&		&		&		&		&		&		&		&		&		&		&		&		&	*	&	*	&		&		&		&		&		&		&		&		&		&		&		&		&		&		&	*	&		&		&		&		&		&		&		\\
\hline
53	&	\cite{ling2012robust}	&		&		&		&		&	*	&		&		&		&		&		&		&		&		&		&		&	*	&		&		&		&	*	&		&		&		&		&		&	*	&		&		&		&		&		&		&	*	&		&		&		&		\\
\hline
54	&	\cite{pflug20121}	&	*	&		&		&		&		&		&	*	&		&		&		&		&		&		&		&		&		&	*	&		&		&		&		&		&	*	&		&		&		&	*	&		&		&		&	*	&	*	&		&		&		&		&		\\
\hline
55	&	\cite{lim2012robust}	&		&		&		&		&		&		&		&		&		&	*	&		&		&		&		&		&		&		&	*	&		&		&		&		&		&		&	*	&		&		&	*	&		&		&		&		&		&		&		&		&		\\
\hline
56	&	\cite{chen2012robust}	&		&		&		&		&		&		&		&		&		&		&	*	&		&		&		&		&		&		&		&	*	&	*	&		&		&		&		&		&	*	&		&		&		&		&		&		&		&		&	*	&		&		\\
\hline
57	&	\cite{nguyen2012robust}	&	*	&		&		&		&		&		&		&		&		&		&		&		&		&		&		&		&		&		&	*	&		&		&		&		&	*	&		&	*	&		&		&		&		&		&		&		&		&	*	&		&		\\
\hline
58	&	\cite{hauser2013relative}	&	*	&		&		&		&		&		&		&		&		&		&		&		&		&	*	&	*	&		&		&		&		&	*	&		&		&		&		&		&		&		&	*	&		&		&		&		&	*	&		&		&		&		\\
\hline
59	&	\cite{kim2013robust}	&	*	&		&		&		&		&		&		&		&		&		&		&		&		&	*	&		&		&		&		&		&	*	&		&		&		&		&		&	*	&		&		&		&		&		&	*	&	*	&		&		&		&		\\
\hline
60	&	\cite{kim2013composition}	&	*	&		&		&		&		&		&		&		&		&		&		&		&		&	*	&		&		&		&		&		&	*	&		&		&		&		&		&	*	&		&		&		&		&		&	*	&	*	&		&		&		&		\\
\hline
61	&	\cite{gulpinar2013robust}	&		&		&		&		&		&		&		&		&	*	&		&		&		&		&	*	&		&		&		&		&		&	*	&		&		&		&		&		&	*	&		&		&		&		&		&		&	*	&		&		&		&		\\
\hline
62	&	\cite{glasserman2013robust}	&		&		&		&		&	*	&		&		&	*	&		&		&		&		&		&		&		&	*	&		&		&		&		&		&		&		&		&		&		&		&		&		&	*	&		&		&		&		&		&		&		\\
\hline
63	&	\cite{deng2013robust}	&		&		&		&		&		&		&	*	&		&		&		&		&	*	&		&	*	&		&		&		&		&		&		&		&		&		&		&	*	&	*	&		&		&		&		&		&	*	&		&		&		&		&		\\
\hline
64	&	\cite{ghahtarani2013robust}	&		&		&		&		&		&		&		&		&		&		&		&	*	&		&		&		&		&		&		&	*	&	*	&		&		&		&		&		&	*	&		&		&		&		&	*	&		&		&		&		&		&		\\
\hline
65	&	\cite{fliege2014robust}	&	*	&		&		&		&		&		&		&		&		&		&		&		&		&	*	&	*	&		&		&		&		&	*	&		&		&		&		&		&	*	&		&		&		&		&		&		&	*	&		&		&		&		\\
\hline
66	&	\cite{pinar2014mean}	&	*	&		&		&		&		&		&		&		&		&		&		&		&		&	*	&		&		&	*	&		&		&	*	&		&	*	&		&		&		&	*	&	*	&		&		&		&		&		&	*	&		&		&		&		\\
\hline
67	&	\cite{kim2014deciphering}	&	*	&		&		&		&		&		&		&		&		&		&		&		&		&	*	&		&		&		&		&		&	*	&		&		&		&		&		&	*	&		&		&		&		&		&		&	*	&		&		&		&		\\
\hline
68	&	\cite{kim2014robust}	&		&		&		&		&	*	&		&		&		&		&		&		&		&		&		&		&	*	&		&		&		&	*	&		&		&		&		&		&	*	&		&		&		&		&		&		&	*	&		&		&		&		\\
\hline
69	&	\cite{recchia2014robust}	&		&		&		&		&		&		&		&		&		&		&		&		&		&	*	&		&		&		&		&		&	*	&		&		&		&		&		&	*	&		&		&		&		&	*	&		&	*	&		&		&		&		\\
\hline
70	&	\cite{kakouris2014robust}	&		&		&		&		&		&		&	*	&		&		&		&		&		&		&		&		&		&	*	&		&		&		&		&		&		&		&	*	&		&	*	&		&		&		&		&		&		&		&		&	*	&		\\
\hline
71	&	\cite{han2017dynamic}	&		&		&		&		&		&		&	*	&		&		&		&		&		&		&		&		&		&	*	&		&		&		&		&		&		&		&	*	&		&	*	&		&		&		&		&		&		&		&		&	*	&		\\
\hline
72	&	\cite{zhu2014portfolio}	&		&		&		&		&		&		&	*	&		&		&		&		&		&		&		&		&		&	*	&		&	*	&	*	&		&		&		&		&		&	*	&	*	&		&		&		&	*	&		&	*	&		&		&		&		\\
\hline
73	&	\cite{liu2014regime}	&		&		&		&		&		&		&	*	&	*	&		&		&		&		&		&		&		&		&	*	&		&		&		&		&	*	&		&		&		&		&	*	&		&		&		&		&		&	*	&		&		&		&		\\
\hline
74	&	\cite{flor2014robust}	&		&		&		&		&		&		&		&		&		&		&		&		&	*	&	*	&		&		&		&		&		&		&		&		&		&		&		&		&		&		&		&	*	&		&		&		&		&		&		&		\\
\hline
75	&	\cite{pae2014log}	&		&		&		&		&		&		&		&		&		&	*	&		&		&		&		&		&		&		&	*	&		&	*	&		&		&		&		&		&	*	&		&		&		&		&	*	&		&		&		&		&		&		\\
\hline
76	&	\cite{gulpiotanar2014robust}	&		&		&		&		&		&		&		&		&		&	*	&		&		&	*	&		&		&		&		&		&	*	&	*	&	*	&		&		&		&		&	*	&		&		&		&		&		&		&	*	&		&		&		&		\\
\hline
77	&	\cite{ling2014robust}	&		&		&		&		&		&		&		&		&		&		&	*	&		&		&		&		&		&	*	&		&		&	*	&		&	*	&		&		&		&		&	*	&		&		&		&		&		&	*	&		&		&		&		\\
\hline
78	&	\cite{kapsos2014worst}	&		&		&		&		&		&		&		&		&		&		&		&	*	&		&		&		&		&	*	&		&		&	*	&		&		&		&	*	&		&		&	*	&		&		&		&	*	&		&	*	&		&		&		&		\\
\hline
79	&	\cite{MAILLET2015289}	&		&	*	&		&		&		&		&		&		&		&		&		&		&		&		&	*	&		&		&		&		&		&		&		&		&		&	*	&	*	&		&		&		&		&		&		&	*	&		&		&		&		\\
\hline
80	&	\cite{kim2015focusing}	&	*	&		&		&		&		&		&		&		&		&		&		&		&		&	*	&		&		&		&		&		&	*	&		&		&		&		&		&	*	&		&		&		&		&		&	*	&	*	&		&		&		&		\\
\hline
81	&	\cite{doan2015robustness}	&		&		&		&		&		&		&	*	&		&		&		&		&		&		&		&		&		&	*	&		&		&		&		&		&		&	*	&		&		&	*	&		&		&		&	*	&		&		&		&		&		&		\\
\hline
82	&	\cite{rezaie2015ideal}	&		&		&		&		&		&		&	*	&		&		&		&		&		&		&	*	&		&		&		&		&		&	*	&		&		&		&		&		&	*	&		&		&		&		&	*	&		&		&		&		&		&		\\
\hline
83	&	\cite{marzban2015developing}	&		&		&		&		&		&		&		&	*	&		&		&		&		&		&	*	&		&		&		&		&		&	*	&		&		&		&		&		&	*	&		&		&		&		&	*	&		&		&		&		&		&		\\
\hline
84	&	\cite{liu2015robust}	&		&		&		&		&		&	*	&		&	*	&		&		&		&		&		&	*	&		&		&		&		&		&	*	&		&		&		&		&		&	*	&		&		&		&		&		&	*	&		&		&		&		&		\\
\hline
85	&	\cite{desmettre2015robust}	&		&		&		&		&		&		&		&	*	&		&		&		&		&		&		&		&		&		&		&	*	&		&		&		&		&		&	*	&	*	&		&		&		&		&	*	&		&		&		&		&		&		\\
\hline
86	&	\cite{pinar2016robust}	&	*	&		&		&		&		&		&		&		&		&		&		&		&		&	*	&		&		&		&		&		&	*	&		&		&		&		&		&	*	&		&		&		&		&		&		&	*	&		&		&		&		\\
\hline
87	&	\cite{li2016portfolio}	&		&		&	*	&		&		&		&		&		&		&		&		&		&		&	*	&		&		&		&		&		&		&	*	&		&		&		&		&	*	&		&		&		&		&		&		&	*	&		&		&		&		\\
\hline
88	&	\cite{rujeerapaiboon2016robust}	&		&		&		&		&		&		&	*	&		&		&		&		&		&		&		&		&		&	*	&		&		&		&		&	*	&		&		&		&		&	*	&		&		&		&		&		&		&	*	&		&		&		\\
\hline
89	&	\cite{fernandes2016adaptive}	&		&		&		&		&		&		&		&	*	&		&		&		&		&		&	*	&		&		&		&		&		&		&		&		&		&		&	*	&	*	&		&		&		&		&	*	&		&		&		&		&		&		\\
\hline
90	&	\cite{yu2016regime}	&		&		&		&		&		&		&	*	&	*	&		&		&		&		&		&	*	&		&		&	*	&		&		&	*	&		&		&		&		&		&	*	&		&		&		&		&		&		&		&		&		&		&		\\
\hline
91	&	\cite{gulpinar2016robust}	&		&		&		&		&		&		&		&		&	*	&		&		&		&		&	*	&		&		&		&		&		&	*	&	*	&		&		&		&		&	*	&		&		&		&		&		&		&	*	&		&		&		&		\\
\hline
92	&	\cite{peykani2016utilizing}	&		&		&		&		&		&		&		&		&		&		&		&		&	*	&	*	&		&		&		&		&		&	*	&		&		&		&		&		&	*	&		&		&		&		&	*	&		&		&		&		&		&		\\
\hline
93	&	\cite{XIDONAS201760}	&		&	*	&		&		&		&		&		&		&		&		&		&		&		&		&	*	&		&		&		&		&		&		&		&		&	*	&		&	*	&		&		&		&		&		&		&		&		&	*	&		&		\\
\hline
94	&	\cite{xidonas2017robust}	&		&		&	*	&		&		&		&		&		&		&		&		&		&		&	*	&		&		&		&		&		&		&		&		&		&	*	&		&		&		&	*	&		&		&		&		&		&		&		&		&	*	\\
\hline
95	&	\cite{belhajjam2017robust}	&	*	&		&		&		&		&		&	*	&		&		&		&		&		&		&		&		&		&	*	&		&		&		&		&	*	&		&		&		&		&	*	&		&		&		&		&	*	&	*	&		&		&		&		\\
\hline
96	&	\cite{yu2017incorporating}	&		&		&		&		&		&		&	*	&		&		&		&		&		&		&		&		&		&	*	&		&		&		&		&		&		&	*	&		&		&		&	*	&		&		&	*	&		&		&		&		&		&		\\
\hline
97	&	\cite{cong2017robust}	&	*	&		&		&		&		&		&		&	*	&		&		&		&		&		&	*	&	*	&		&		&		&		&	*	&		&		&		&		&		&	*	&		&		&		&		&		&	*	&		&		&		&		&		\\
\hline
98	&	\cite{ling2017robust}	&		&		&		&		&		&		&	*	&	*	&		&		&		&		&		&		&		&		&	*	&		&		&		&		&	*	&		&		&		&		&	*	&		&		&		&		&		&		&	*	&		&		&		\\
\hline
99	&	cite{platanakis2017asset}	&		&		&		&		&	*	&		&		&		&	*	&		&		&		&		&	*	&		&		&		&		&	*	&	*	&		&		&		&		&		&	*	&		&		&		&		&		&		&	*	&		&		&		&		\\
\hline
100	&	\cite{bo2017robust}	&		&		&		&		&		&		&		&		&		&		&		&		&	*	&		&		&		&		&		&		&		&		&		&		&		&		&		&		&		&		&	*	&		&		&		&		&		&		&		\\
\hline
101	&	\cite{kawas2017log}	&		&		&		&		&		&		&		&		&		&	*	&		&		&		&		&		&		&		&	*	&		&	*	&		&		&		&		&		&	*	&	*	&		&		&		&	*	&		&		&		&		&		&		\\
\hline
102	&	\cite{gulpinar2017robust}	&		&		&		&		&		&		&	*	&		&		&		&		&		&		&		&		&		&		&		&	*	&	*	&	*	&		&		&		&		&	*	&		&		&		&		&		&		&	*	&		&		&		&		\\
\hline
103	&	\cite{sharma2017omega}	&		&		&		&		&		&		&	*	&		&		&		&		&	*	&		&		&		&		&	*	&		&		&	*	&		&		&		&	*	&		&		&	*	&		&		&		&	*	&		&	*	&		&		&		&		\\
\hline
104	&	\cite{ding2018robust}	&	*	&		&		&		&		&		&		&		&		&		&		&		&		&	*	&	*	&		&		&		&		&		&		&		&	*	&		&		&		&	*	&		&		&		&		&		&		&		&		&	*	&		\\
\hline
105	&	\cite{article}	&	*	&		&		&		&		&		&		&		&		&		&		&		&		&	*	&	*	&		&		&		&		&	*	&		&		&		&		&		&		&		&	*	&		&		&		&		&	*	&		&		&		&		\\
\hline
106	&	\cite{kapsos2018robust}	&		&	*	&		&		&		&		&		&		&		&		&		&		&		&		&	*	&		&		&		&		&	*	&		&		&		&	*	&		&	*	&		&		&		&		&		&	*	&		&		&		&		&		\\
\hline
107	&	\cite{chen2018robust}	&	*	&		&		&		&		&		&		&		&		&		&		&		&		&	*	&	*	&		&		&		&		&	*	&		&		&		&		&		&	*	&		&		&		&	*	&		&		&	*	&		&		&		&		\\
\hline
108	&	\cite{ghahtarani2018robust}	&		&		&	*	&		&		&		&		&		&		&		&		&		&		&	*	&		&		&		&		&		&	*	&		&		&		&		&		&	*	&		&		&		&		&	*	&		&		&		&		&		&		\\
\hline
109	&	\cite{hasuike2018investor}	&		&		&		&		&	*	&		&	*	&		&		&		&		&		&		&		&		&		&	*	&		&		&		&		&	*	&		&		&		&		&	*	&		&		&		&		&	*	&		&		&		&		&		\\
\hline
110	&	\cite{ghahtarani2018development}	&		&		&		&		&		&		&	*	&		&		&		&		&		&		&		&		&		&	*	&		&		&		&		&	*	&		&		&		&		&	*	&		&		&		&	*	&		&		&		&		&		&		\\
\hline
111	&	\cite{pacc2018robust}	&	*	&		&		&		&		&		&	*	&		&		&		&		&		&		&		&		&		&	*	&		&		&		&		&		&	*	&		&		&		&	*	&		&		&		&	*	&	*	&		&		&		&		&		\\
\hline
112	&	\cite{liu2018time}	&		&		&		&		&		&		&	*	&	*	&		&		&		&		&		&		&		&		&	*	&		&		&		&		&	*	&		&		&		&		&	*	&		&		&		&		&		&	*	&		&		&		&		\\
\hline
113	&	\cite{gabrel2018portfolio}	&		&		&		&		&		&		&	*	&		&		&		&		&		&		&	*	&		&		&		&		&		&		&		&		&		&	*	&		&		&		&		&		&	*	&		&		&		&		&	*	&		&		\\
\hline
114	&	\cite{lu2019asset}	&	*	&		&		&		&		&		&		&		&		&		&		&		&		&	*	&	*	&		&		&		&		&	*	&		&		&		&		&		&	*	&		&		&		&		&		&		&	*	&		&		&		&		\\
\hline
115	&	\cite{chen2019robust}	&	*	&		&		&		&		&		&		&		&		&		&		&		&		&	*	&	*	&		&		&		&		&	*	&		&		&		&		&		&	*	&		&		&		&	*	&		&		&	*	&		&		&		&		\\
\hline
116	&	\cite{bai2019improving}	&	*	&		&		&		&		&		&		&		&		&		&		&		&		&	*	&	*	&		&		&		&		&	*	&		&		&		&		&		&	*	&		&		&		&	*	&		&		&	*	&		&		&		&		\\
\hline
117	&	\cite{dai2019sparse}	&	*	&		&		&		&		&		&		&		&		&		&		&		&		&	*	&		&		&		&		&		&	*	&		&		&		&		&		&	*	&		&		&		&		&		&	*	&	*	&		&		&		&		\\
\hline
118	&	\cite{plachel2019unified}	&		&	*	&		&		&		&		&		&		&		&		&		&		&		&		&	*	&		&		&		&		&	*	&		&		&		&		&		&	*	&		&		&		&		&		&	*	&		&		&		&		&		\\
\hline
119	&	\cite{liu2019closed}	&		&		&		&		&		&		&	*	&	*	&		&		&		&		&		&		&		&		&	*	&		&		&		&		&	*	&		&		&		&		&	*	&		&	*	&		&		&		&		&		&		&		&	*	\\
\hline
120	&	\cite{kang2019data}	&		&		&		&		&		&		&	*	&		&		&		&		&		&		&		&		&		&	*	&		&		&		&		&	*	&		&		&		&		&	*	&		&		&		&		&		&	*	&		&		&		&		\\
\hline
121	&	\cite{goel2019robust}	&		&		&		&		&		&		&	*	&		&		&		&		&		&		&		&		&		&	*	&		&		&		&		&		&		&		&	*	&		&	*	&		&		&		&	*	&		&		&		&		&		&		\\
\hline
122	&	\cite{kara2019stability}	&		&		&		&		&		&		&	*	&		&		&		&		&		&		&	*	&		&		&		&		&		&		&		&		&		&		&	*	&	*	&		&		&		&		&	*	&		&		&		&		&		&		\\
\hline
123	&	\cite{ling2019robust}	&		&		&		&	*	&	*	&		&		&	*	&		&		&		&		&		&	*	&		&		&		&		&		&		&	*	&		&		&		&		&	*	&		&		&		&		&		&		&	*	&		&		&		&		\\
\hline
124	&	\cite{yu2019realized}	&		&		&		&		&		&		&	*	&		&		&		&		&	*	&		&		&		&		&	*	&		&		&	*	&		&		&		&	*	&		&		&	*	&		&		&		&	*	&		&		&		&		&		&		\\
\hline
125	&	\cite{peykani2019stock}	&		&		&		&		&		&		&		&		&		&		&		&		&	*	&		&		&		&		&		&		&		&		&		&		&	*	&		&	*	&		&		&		&		&	*	&		&		&		&		&		&		\\
\hline
126	&	\cite{khodamoradi2020robust}	&	*	&		&		&		&		&		&		&		&		&		&		&		&		&	*	&	*	&		&		&		&		&	*	&		&		&		&		&		&	*	&		&		&		&		&		&		&		&		&	*	&		&		\\
\hline
127	&	\cite{lee2020sparse}	&	*	&		&		&		&		&		&		&		&		&		&		&		&		&	*	&		&		&		&		&		&	*	&		&		&		&		&		&	*	&		&		&		&		&		&		&	*	&		&		&		&		\\
\hline
128	&	\cite{peykani2020novel}	&	*	&		&	*	&		&		&		&		&		&		&		&		&		&		&	*	&		&		&		&		&		&	*	&		&		&		&		&		&	*	&		&		&		&		&	*	&		&		&		&		&		&		\\
\hline
129	&	\cite{georgantas2021robust}	&	*	&		&		&		&		&		&	*	&		&		&		&		&	*	&		&	*	&		&		&	*	&		&		&	*	&		&		&		&		&		&	*	&	*	&		&		&		&	*	&		&		&		&		&		&		\\
\hline
130	&	\cite{kouaissah2021robust}	&		&		&		&		&		&		&		&		&		&		&		&		&	*	&	*	&		&		&		&		&		&	*	&		&		&		&		&		&	*	&		&		&		&		&		&		&	*	&		&		&		&		\\
\hline
131	&	\cite{chakrabarti2021parameter}	&		&	*	&		&		&		&		&		&		&		&		&		&	*	&		&	*	&	*	&		&		&		&		&	*	&		&		&		&		&		&	*	&		&	*	&		&		&	*	&	*	&		&		&		&		&		\\
\hline
132	&	\cite{benati2021relative}	&		&		&		&		&		&		&	*	&		&		&		&		&		&		&	*	&		&		&		&		&		&		&		&		&		&	*	&		&		&		&	*	&		&		&		&		&		&		&		&		&	*	\\
\hline
133	&	\cite{ashrafi2021study}	&		&		&		&		&		&		&		&		&		&		&		&		&	*	&	*	&		&		&		&		&		&	*	&		&		&		&		&		&	*	&		&		&		&		&	*	&		&		&		&		&		&		\\
\hline
134	&	\cite{jiang2021robust}	&	*	&		&		&		&		&		&		&	*	&		&		&		&		&		&	*	&		&		&		&		&		&	*	&		&		&		&		&		&	*	&		&		&		&		&		&		&	*	&		&		&		&		\\
\hline
135	&	\cite{yin2021practical}	&	*	&		&		&		&		&		&		&		&		&		&		&		&		&	*	&		&		&		&		&		&	*	&		&		&		&		&		&	*	&		&		&		&		&		&	*	&	*	&		&		&		&		\\
\hline
136	&	\cite{baviera2021model}	&	*	&		&		&		&		&		&		&		&		&		&		&		&		&		&		&		&	*	&		&		&		&		&		&	*	&		&		&		&	*	&		&		&		&		&	*	&		&		&		&		&		\\
\hline
137	&	\cite{huang2021sparse}	&		&		&		&		&		&		&	*	&		&		&		&		&		&		&	*	&		&		&		&		&		&		&		&	*	&		&		&		&		&	*	&		&		&		&	*	&		&		&		&		&		&		\\
\hline
138	&	\cite{sehgal2021robust}	&		&		&	*	&		&		&		&	*	&		&		&		&		&	*	&		&	*	&		&		&		&		&		&	*	&		&		&		&		&		&	*	&		&		&		&		&	*	&		&		&		&		&		&		\\
\hline
139	&	\cite{caccador2021portfolio}	&	*	&	*	&		&		&		&		&		&		&		&		&		&		&		&	*	&		&		&		&		&		&		&		&		&		&	*	&		&		&		&	*	&		&		&		&		&		&		&		&		&	*	\\
\hline
140	&	\cite{gajek2021robust}	&		&		&		&		&		&		&	*	&		&	*	&		&		&		&		&		&		&		&	*	&		&		&		&		&		&		&	*	&		&		&	*	&		&		&		&		&	*	&		&		&		&		&		\\
\hline
141	&	\cite{zhao2021robust}	&		&		&		&		&		&		&	*	&		&		&		&		&		&		&		&		&		&	*	&		&		&		&		&	*	&		&		&		&		&	*	&		&		&		&		&	*	&		&		&		&		&		\\
\hline
142	&	\cite{swain2021robust}	&	*	&	*	&		&		&		&		&		&		&		&		&		&		&		&	*	&	*	&		&		&		&		&	*	&		&		&		&		&		&	*	&		&		&		&		&		&	*	&		&		&		&		&		\\\hline
\end{longtable}
}
\end{center}
%\end{landscape}

Moreover, figures provide some statistics about the reviewed papers. 

\begin{figure}[H]
\centering
%\begin{multicols}
\includegraphics[width=120mm]{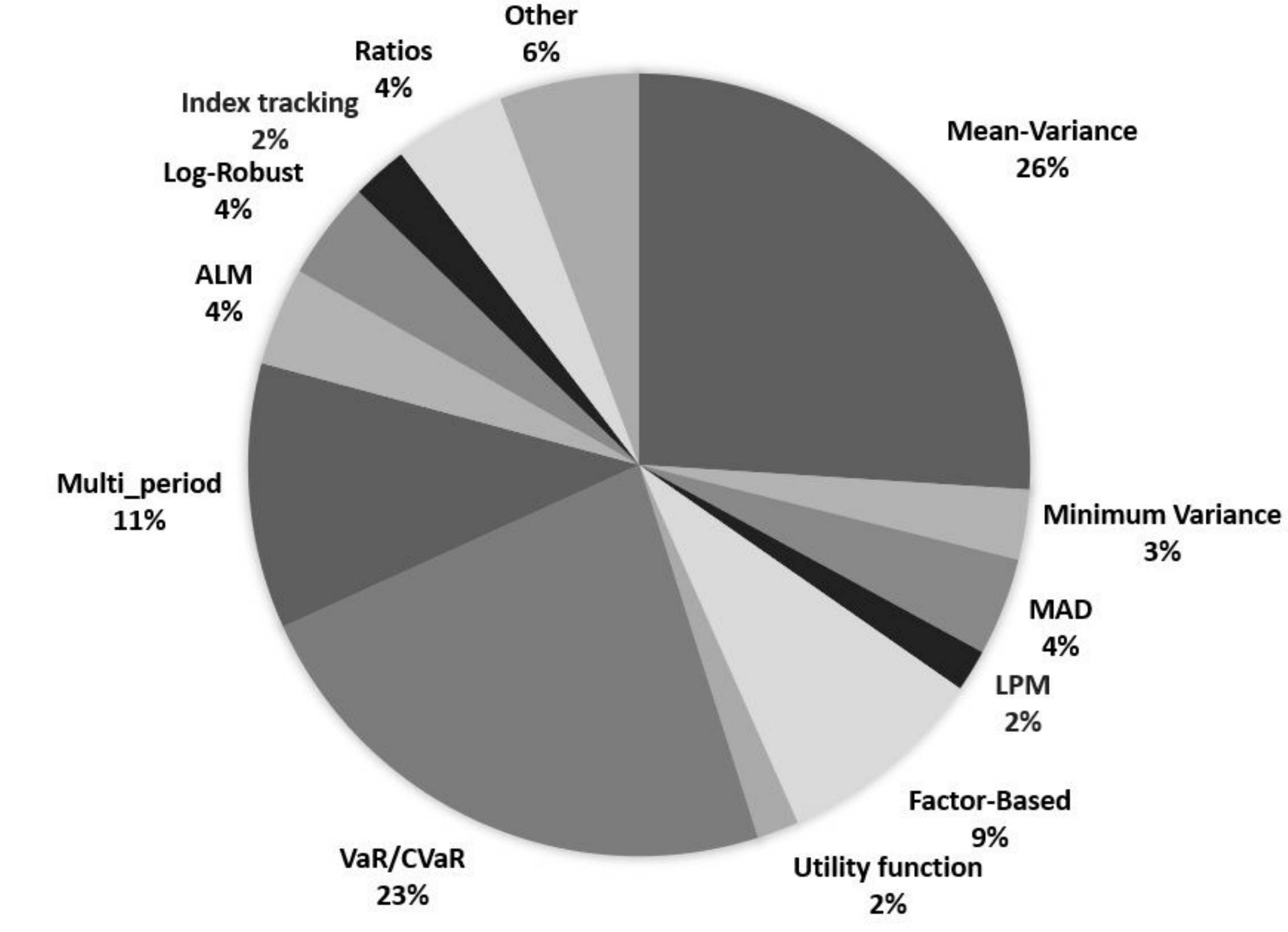}
%\end{multicols}
\caption{Distribution of articles based on PSP type}
\label{fig:4-1}
\end{figure}

\begin{figure}[H]
\centering
%\begin{multicols}
\includegraphics[width=120mm]{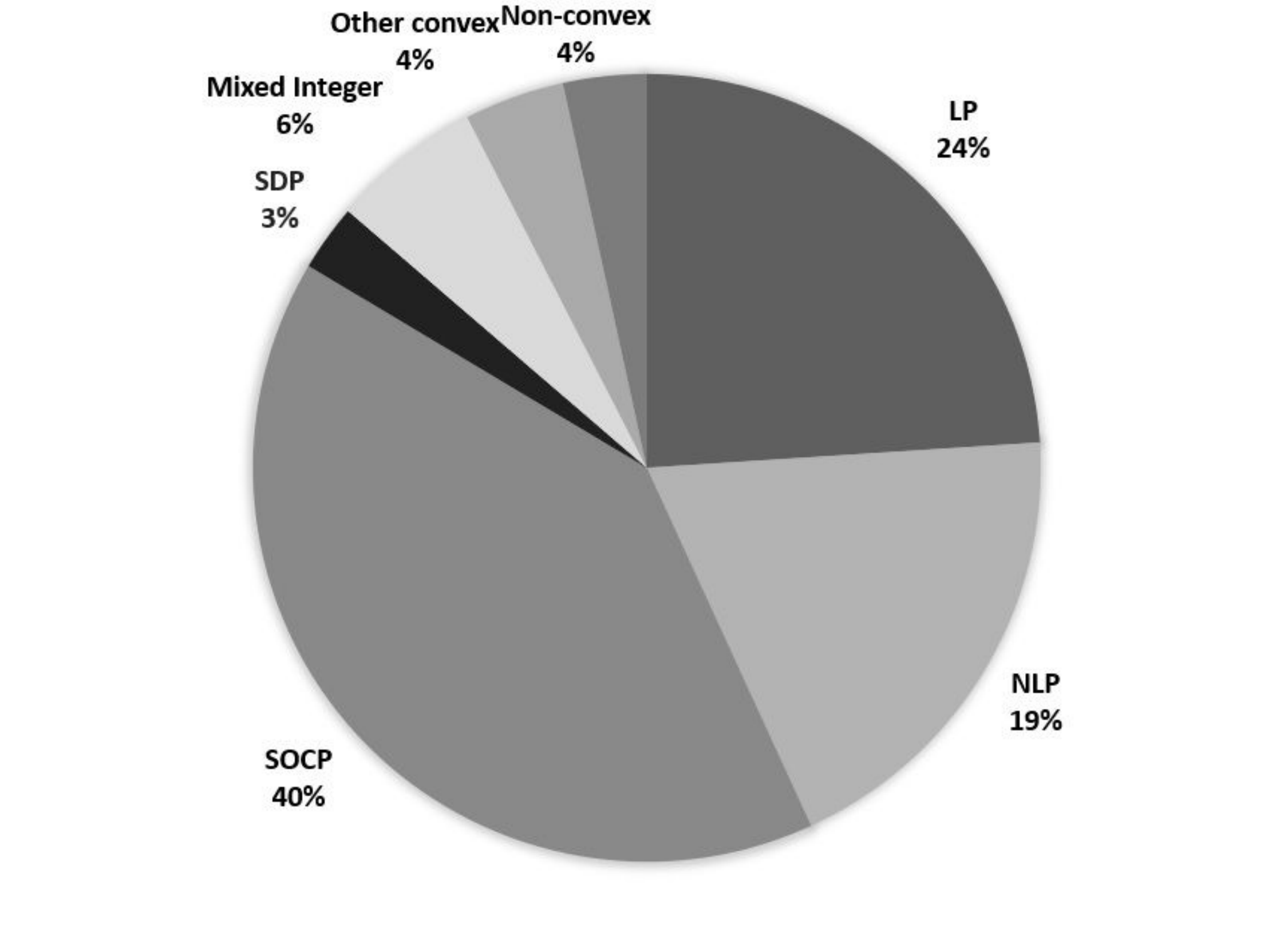}
%\end{multicols}
\caption{Distribution of articles based on optimization problem types}
\label{fig:8-1}
\end{figure}

\begin{figure}[H]
\centering
%\begin{multicols}
\includegraphics[width=120mm]{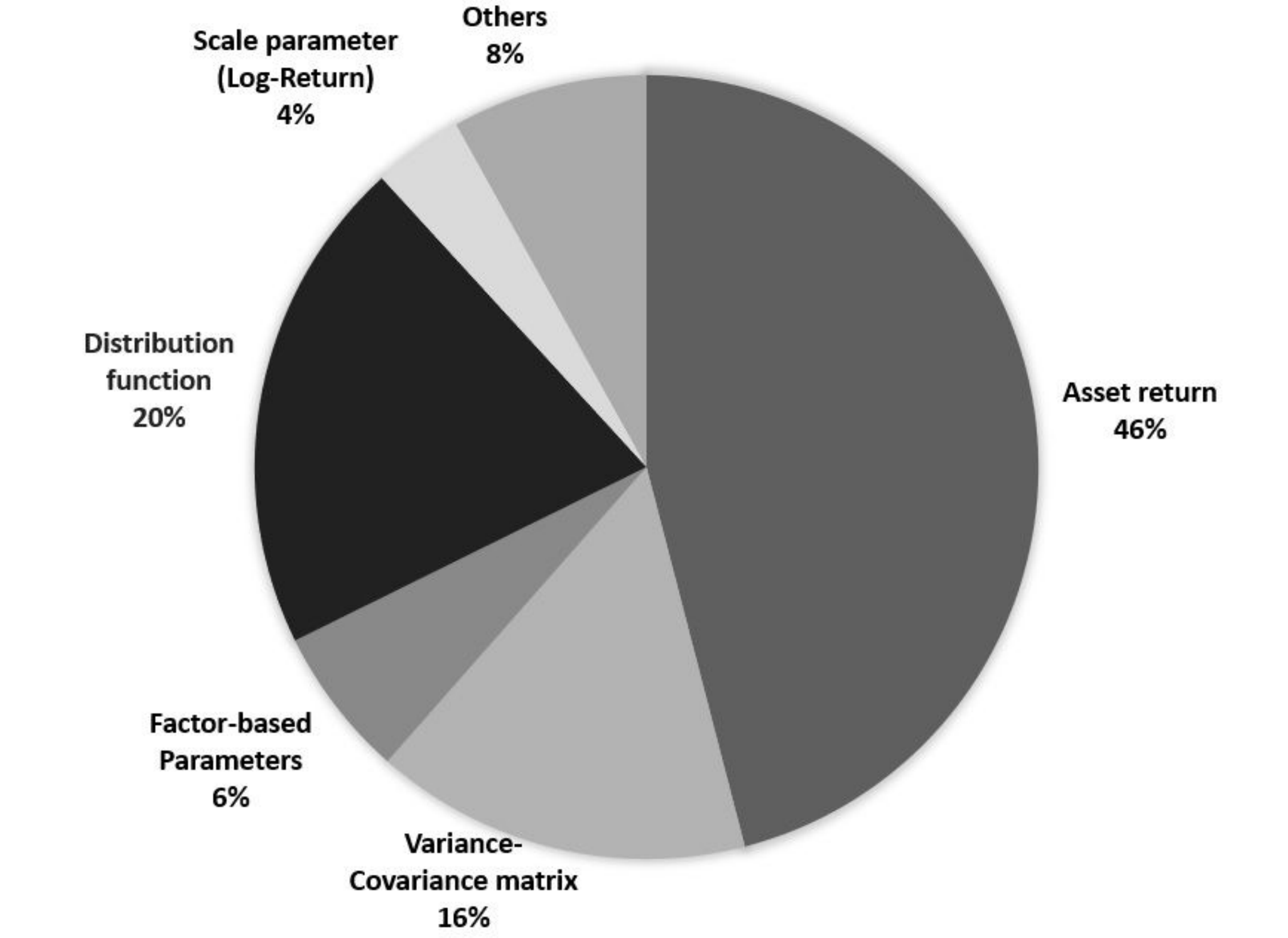}
%\end{multicols}
\caption{Distribution of articles based on uncertain parameters}
\label{fig:5-1}
\end{figure}

\begin{figure}[H]
\centering
%\begin{multicols}
\includegraphics[width=120mm]{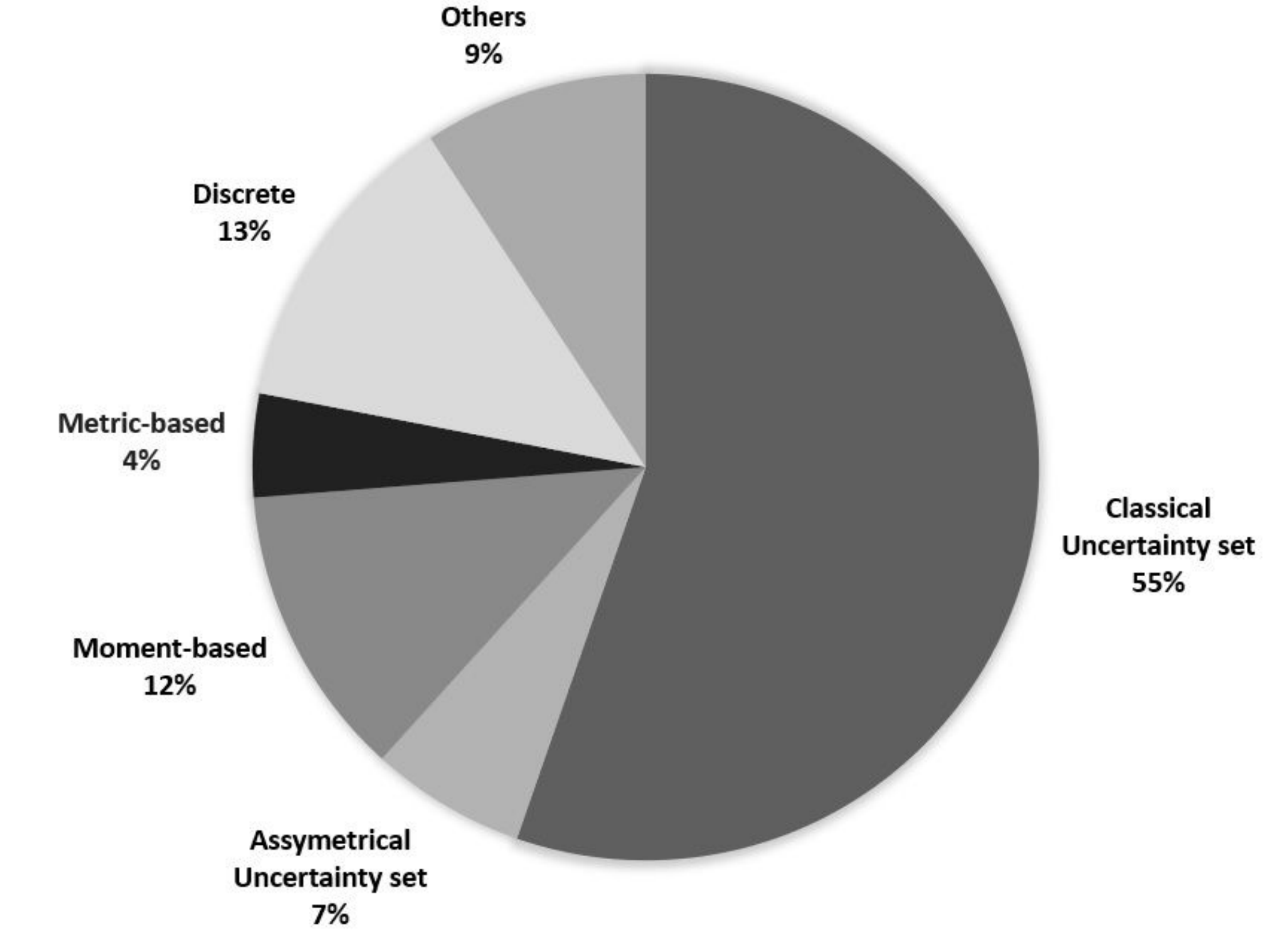}
%\end{multicols}
\caption{Distribution of articles based on uncertainty sets}
\label{fig:6-1}
\end{figure}

\begin{figure}[H]
\centering
%\begin{multicols}
\includegraphics[width=120mm]{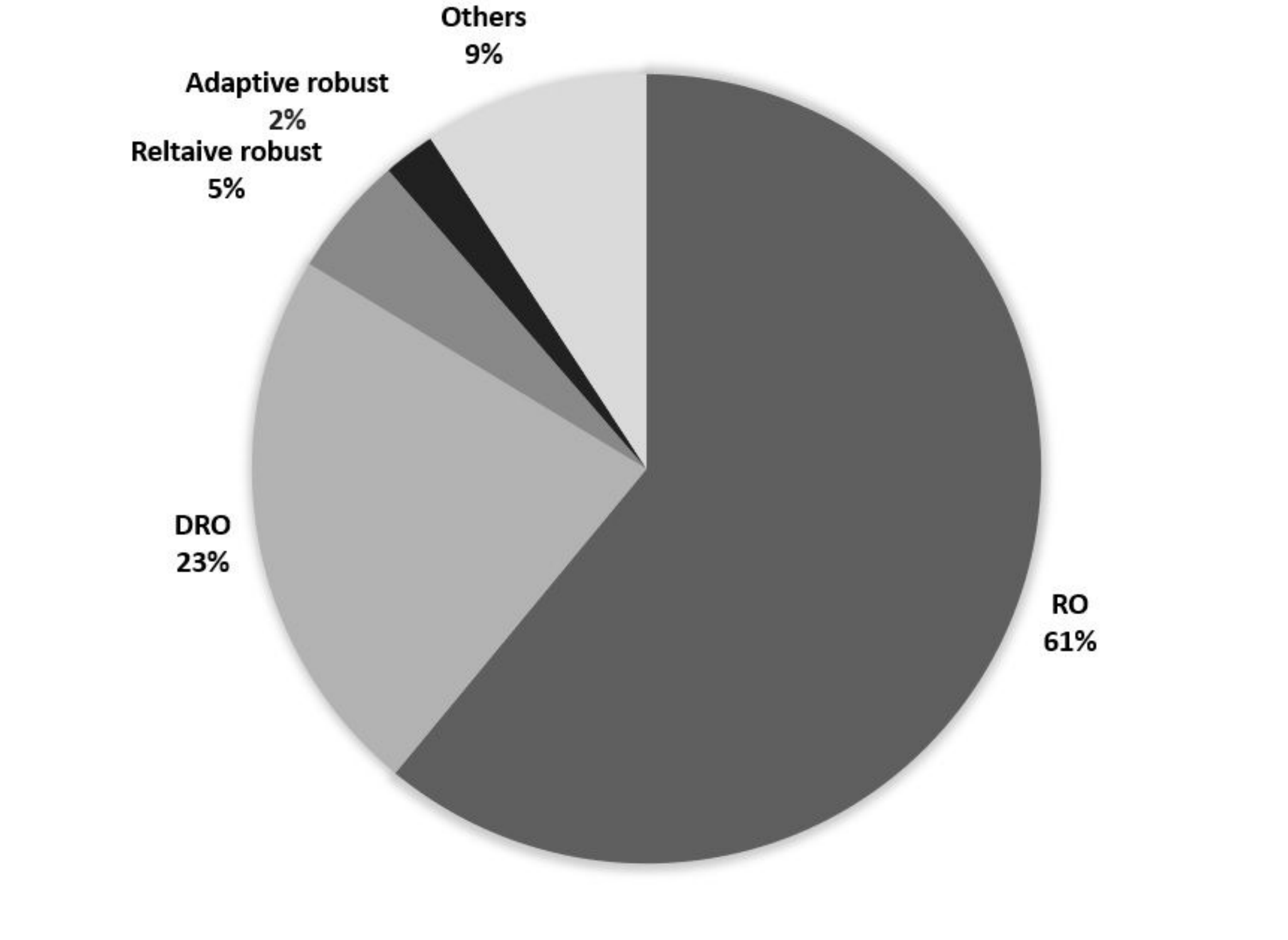}
%\end{multicols}
\caption{Distribution of articles based on RO methods}
\label{fig:7-1}
\end{figure}

Figures \ref{fig:4-1}, \ref{fig:8-1} show that \(\%26\) of published articles used mean-variance and \(\%23\) of published articles proposed robust VaR/CVaR formulations. Moreover, majority of robust PSPs leads to SOCP, and LP. The third robust counterpart type is NLP with \(\%19\). Figures \ref{fig:5-1}, \ref{fig:6-1} show that \(\%46\) of published articles consider asset return as uncertain parameters. Another important classification of articles is based on type of uncertainty set. This figure also shows the distribution of uncertainty sets in published articles. It demonstrates that about \(\%55\) of articles use classical uncertainty sets include box, ellipsoidal, budgeted, and polyhedral. Figure \ref{fig:7-1} illustrates the distribution of RO methods in published articles. This figure shows that mostly classical RO and DRO are used in articles.

\section{Conclusions and Future Research Directions}\label{conclusions}
Portfolio selection has been a fertile area for applying modern RO techniques as evident by the large number of robust PSP articles published in the last two decades. The inherent uncertainty about future asset returns, the abundance of public data available and the risk-averse nature of most investors make RO an appealing approach in this area. As shown in this review paper, a wide range of robust PSP variants was studied, from a ``plain vanilla" single-period, mean-variance PSP with a simple box uncertainty set (\eg \cite{tutuncu2004robust}) to formulations that consider advanced risk measures (\eg \cite{ghahtarani2018development}, \cite{huang2010portfolio}), adaptive uncertainty sets (\eg \cite{yu2016regime}), real-life investment strategies (\eg \cite{pflug20121}, \cite{pacc2018robust}) and dynamic portfolio balancing (\eg \cite{ling2019robust}, \cite{cong2017robust}). This variety of modeling assumptions and approaches and the overlaps among them make it difficult to develop a unifying framework for robust PSPs, yet we adopted a multi-dimensional classification scheme that depends on the risk measure to be optimized, the type of uncertain parameters, the approach used to capture uncertainty and the the planning horizon (\ie single- vs. multi-period).

Despite the surge of interest about robust PSPs in the research community, this area has received little attention from practitioners. A possible reason for such a rift between theory and practice is that research in this area was often driven by advancements in operations research methods rather than being in response to the real needs of the financial industry. Moreover, the value of using robust approaches might not be readily apparent to practitioners who are accustomed to classical PSP models. Therefore, experimental studies, like those presented in  \cite{kim2013robust}, \cite{kim2014deciphering}, \cite{kim2013composition}, \cite{kim2018robust}, \cite{kim2015focusing}, \cite{schottle2009robustness}, are crucial for bridging this gap. The fact that tractable reformulations of most robust counterparts are more complex, both conceptually and computationally, than their corresponding deterministic formulations might make robust reformulations less attractive for practitioners (\eg \cite{kouvelis1997robust}, \cite{hauser2013relative}, \cite{article}, \cite{lim2012robust}, \cite{huang2010portfolio}). 

Nevertheless, the perception of robust optimization as an overly conservative portfolio selection approach is probably the major obstacle to its wide adoption by investment professionals. The reader can easily notice that this issue has received a lot of attention in the robust PSP research. Approaches proposed in the literature to attenuate the conservatism of robust formulations include: using controllable uncertainty sets (\eg ellipsoid  \citep{fabozzi2007robust} or budget \citep{liu2015robust}), data-driven approaches (\eg \cite{doan2015robustness}, \cite{bienstock2007histogram}), alternative risk measures (\eg relative log-return \citep{lim2012robust} or \citep{huang2010portfolio}), distributionally robust optimization (\eg \cite{ling2014robust}) and regime-dependent robust models (\eg \cite{liu2014regime}, \cite{yu2016regime}). While these approached can be effective in controlling conservatism and providing well-balanced solutions, they often lead to models that are challenging to handle.

Given that asset returns do not generally behave like independent random variables, but are instead dependent on common factors and have significant temporal correlations, trying to capture the variability of returns directly often leads to large uncertainty sets and hence conservative solutions. Instead, robust factor models deals with the uncertainty in the independent factors themselves, thus lead less conservative formulations. However, as \cite{lu2006new} noted, selecting the suitable factors for the model and adjusting their weights are still worthy of further investigation. Another promising direction is to use \emph{dynamic uncertainty sets}, that incorporate time-series models to capture auto-correlations in asset returns. Dynamic sets have been shown to result in less pessimistic solutions compared to static ones in other applications \citep{lorca2014adaptive, lorca2016multistage}. 

An important advantage of financial markets is the abundance of historical data that can be used to build uncertainty and ambiguity sets for uncertain parameters. Although data-driven robust formulations for a few variants of the PSP have been proposed in the literature (\eg \cite{bienstock2007histogram}, \cite{kawas2017log}, \cite{rujeerapaiboon2016robust}, \cite{doan2015robustness}, \cite{lotfi2018robust}, \cite{liu2019closed}, \cite{kang2019data}), this is still a promising area for future research given the recently-proposed techniques for constructing and sizing uncertainty sets to achieve desirable properties (see \eg \cite{bertsimas2009constructing}, \cite{bertsimas2018data}). In a related matter, and as noted by \cite{kang2019data} in the context of robust CVaR optimization, it is still unclear which ambiguity sets should be used for DRO PSPs and how they should be sized to provide the best out-of-sample performance. With the plethora of ambiguity set structures proposed in recent year, investigating new variants of the distributionally robust PSPs is a plausible research direction.

Another promising research direction is the application of ``soft" robust optimization methods to financial problems. A drawback of classical robust optimization is that it tries to capture most possible realizations of the parameters within the uncertainty set, which usually results in large sets and conservative solutions. Alternatively, one can construct smaller uncertainty sets that include only a subset of these possible realization and allow robust constraints to be violated, yet with penalties. Examples of these approaches include Globalized Robust Optimization \citep{ben2017globalized}, Robustness Optimization \citep{long2019dao} and Almost Robust Optimization \citep{baron2019almost}. Soft robust optimization methods are still scarcely applied in the PSP literature (see \cite{recchia2014robust}), but have the potential for providing a trade-off between robustness and the quality of solutions.

\bibliographystyle{apa-good}
\bibliography{main}

\end{document}